\documentclass[final,5p,times,twocolumn]{elsarticle}
\usepackage{booktabs} 
\usepackage{amsmath}
\usepackage{url} 
\usepackage{subcaption}
\usepackage{pdfpages}
\usepackage{threeparttable} 
\usepackage{colortbl}
\usepackage{multirow}
\usepackage{xcolor}
\usepackage{graphicx}
\graphicspath{ {figures/} }
\usepackage{pgfplots}
\usepackage{pgfplotstable}
\usepgfplotslibrary{groupplots}
\usepgfplotslibrary{polar}
\usepackage{hyperref}
\pgfplotstableset{
    /color cells/min/.initial=0,
    /color cells/max/.initial=1000,
    /color cells/textcolor/.initial=,
    color cells/.code={%
        \pgfqkeys{/color cells}{#1}%
        \pgfkeysalso{%
            postproc cell content/.code={%
                \begingroup
                \pgfkeysgetvalue{/pgfplots/table/@preprocessed cell content}\value
\ifx\value\empty
\endgroup
\else
                \pgfmathfloatparsenumber{\value}%
                \pgfmathfloattofixed{\pgfmathresult}%
                \let\value=\pgfmathresult
                \pgfplotscolormapaccess
                    [\pgfkeysvalueof{/color cells/min}:\pgfkeysvalueof{/color cells/max}]%
                    {\value}%
                    {\pgfkeysvalueof{/pgfplots/colormap name}}%
                \pgfkeysgetvalue{/pgfplots/table/@cell content}\typesetvalue
                \pgfkeysgetvalue{/color cells/textcolor}\textcolorvalue
                \toks0=\expandafter{\typesetvalue}%
                \xdef\temp{%
                    \noexpand\pgfkeysalso{%
                        @cell content={%
                            \noexpand\cellcolor[rgb]{\pgfmathresult}%
                            \noexpand\definecolor{mapped color}{rgb}{\pgfmathresult}%
                            \ifx\textcolorvalue\empty
                            \else
                                \noexpand\color{\textcolorvalue}%
                            \fi
                            \the\toks0 %
                        }%
                    }%
                }%
                \endgroup
                \temp
\fi
            }%
        }%
    }
}

\journal{Computer and Security}

\begin{document}

\begin{frontmatter}

\title{Does Travel Stage Matter? How Leisure Travellers Perceive Their Privacy Attitudes Towards Personal Data Sharing Before, During, and After Travel}

\author[inst1]{Haiyue Yuan, Shujun Li}

\affiliation[inst1]{organization={Institute of Cyber Security for Society (iCSS), School of Computing, University of Kent},
            addressline={University of Kent}, 
            city={Canterbury},
            postcode={CT2 7NP}, 
            country={United Kingdom}}

\author[inst2]{Fatima Gillani}

\affiliation[inst2]{organization={Aston Business School, Aston University},
            addressline={Aston University}, 
            city={Birmingham},
            postcode={B4 7ET}, 
            country={United Kingdom}}

\author[inst3]{Xiao Ma}

\affiliation[inst3]{organization={Centre for Business and Industry Transformation (CBIT), Nottingham Business School, Nottingham Trent University},
            addressline={50 Shakespeare Street}, 
            city={Nottingham},
            postcode={NG1 4FQ}, 
            country={United Kingdom}}
            
\begin{abstract}
People's attitudes towards personal data sharing have been extensively researched, however, limited research studied their evolving nature in different contexts, e.g., across different stages of a leisure trip. This paper addresses this gap by exploring how leisure travellers' attitudes towards sharing personal data change across three stages -- before, during and after travel. Analysing data from an online survey with 318 participants, we found that participants' privacy attitudes towards sharing different personal data vary based on sharing purposes and travel stages. Interestingly, participants exhibited a more relaxed attitude towards sharing commonly sensitive personal data (e.g., name, gender, date of birth) compared to other types of personal data. This is likely because sharing such data for travel bookings has become essential, normal, and widely accepted among travellers when using booking sites, which is in line with previous work stating that information easily obtainable is typically not seen as highly confidential. Moreover, our findings revealed that, despite participants' self-reported frequent use of social media platforms, content sharing is minimal on TikTok, YouTube, Snapchat, Pinterest, and Twitter. Conversely, Facebook and Instagram were more commonly used by participants to share travel-related content. This pattern remains consistent across the three stages of travel, suggesting that the stage of travel does not significantly influence how people share on social media platforms, which has been overlooked in past studies. Furthermore, we discovered that a participant's gender, previous travel frequency, and country of residence can influence their perceptions of personal data sharing at different travel stages. This confirms the complex and context-dependent nature of privacy perception and attitudes, which can vary across different situations. Based on the findings observed from this study, we further emphasise and discuss possible applications and potential contributions of our work to the privacy and security community in general.
\end{abstract}

\begin{keyword}
Privacy attitude \sep personal data sharing \sep social media \sep tourism \sep cyber security \sep privacy
\end{keyword}

\end{frontmatter}

\section{Introduction}
\label{sec:introduction}

Technological advancements have rapidly transformed the tourism sector, fundamentally altering how information is presented to travellers, how they plan their travels, make travel bookings, and share personal data with relevant online organisations (e.g., service providers) and other people. The plethora of diverse online platforms, encompassing social networks (e.g., Facebook, Twitter, Instagram, WhatsApp, Snapchat and Pinterest), video sharing services (e.g., TikTok and YouTube), travel-related review sites (e.g., TripAdvisor), and travel booking services (e.g., booking.com and Expedia), have also influenced travellers' attitudes towards privacy. Extensive research has been undertaken to investigate the effects of technological advancement on different stages of leisure travel (i.e., before, during, and after leisure travel). \cite{Fotis-J2012} explored the utilisation of social media platforms during the holiday planning phase and its impact on traveller's decision-making processes and behaviours. Similarly, a more recent work by \cite{Rebeka-Anna-P2022} studied travel decisions, but from the perspective of social media influencers. Furthermore, \cite{Arica-R2022} examined how travellers share their travel experiences on social media platforms and the underlying motivations for sharing personal data. Despite the extensive exploration of privacy concerns and issues related to online services, research specifically focusing on the tourism domain remains limited. \cite{Tussyadiah-I2019} presented a comprehensive review of travellers' privacy concerns and behaviours, highlighting a list of research priorities that address both social and technical aspects of privacy within the tourism context. The study by \cite{Ioannou-A2020} is closer to ours, which looked at travellers' privacy concerns and personal data disclosure. However, their investigation focused on biometric and behavioural data and did not encompass travellers' privacy attitudes or concerns across the full spectrum of the leisure travel journey.

At various stages of travel, leisure travellers' personal data may be collected and shared through mandatory, voluntary, or optional channels for different purposes. For instance, travellers' names and email addresses are required by travel agencies for holiday bookings, whereas dietary requirements are optional. The practices and ability to handle such data by the booking sites can raise privacy concerns and risks. Additionally, travellers may choose to voluntarily share their personal data such as holiday location, pictures, and videos through the use of social media, to their friends, family members or the general public. The widespread use of social media throughout various stages of travel could increase the risk of personal data leakage. Having these in mind, it is a critical consideration within the scope of this research to help understand the landscape of privacy attitudes and behaviour from leisure travellers' perspectives. To the best of our knowledge, there is limited research exploring privacy attitudes and behaviours from these particular perspectives. Although previous studies~\cite{Masur-P2019, Acquisti-A2015} suggested that the privacy levels, privacy perceptions, and cost-benefit trade-offs in privacy decisions are context-dependent, no existing study explores the changes of privacy attitudes and concerns around personal data sharing for different contexts, more specifically, across three stages: before, during, and after travel. To fill the above-mentioned research gap, we conducted a large-scale online survey, aiming to learn about leisure travellers' self-reported privacy attitudes and behaviours about sharing personal data throughout three different travel phases. By exploring and analysing leisure travellers' responses, we aim to answer the following two main research questions (RQs).
\begin{itemize}
\item \textbf{RQ1} \textit{Do leisure travellers have different privacy attitudes towards sharing different types of personal data in different stages of leisure travel?}
\item \textbf{RQ2} How do leisure travellers perceive sharing travel-related content on different social media platforms and are there any observable differences in different stages of leisure travel?
\end{itemize}

The rest of this paper is organised as follows. Section~\ref{sec:relatedwork} reviews related literature. Section~\ref{sec:methodology} describes the experimental design, procedures, and approaches for data collection and analysis. The findings of the study are detailed in Section~\ref{sec:analysis_results}, followed by presenting more discussion about future research and limitations in Section~\ref{sec:discussion}. Finally, Section~\ref{sec:conclusion} provides the conclusion of the paper.

\section{Related work} 
\label{sec:relatedwork}

Travellers today have a wide range of emerging technologies such as mobile apps and AI assistants at their disposal when it comes to planning their leisure travels, making travel bookings, and sharing personal data with travel-related services and individuals. Social media platforms such as Facebook, Twitter, Instagram, WhatsApp, Snapchat, and Pinterest, as well as video-sharing services like TikTok and YouTube, play a significant role in shaping travellers' leisure attitudes and behaviours regarding privacy. Additionally, travel-related review sites such as TripAdvisor and travel booking services like booking.com and Expedia also have a substantial impact. The rapid advancements in technology have facilitated the seamless integration of innovative solutions into every stage of travel, aiming to enhance the overall travel experience. 

During the pre-travel stage, research~\cite{Fotis-J2012} has found evidence that the use of social media can strongly influence changes made to holiday plans before travellers make their final decision. This could largely be attributed to the increased trust in user-generated content over traditional sources such as official tourism websites, travel agencies, and mass media advertising~\cite{Fotis-J2012}. Another study conducted by~\citet{Sangwon-P2017} looked at the use of smartphones for holiday booking at the pre-travel stage. They identified several factors, including time, financial, and privacy/security risks, contributing to the perceived risk that prevents travellers from using their smartphones to book holidays. The study also revealed that personal data collection via smartphones increased the perceived threat and confirmed its impact on travellers' attitudes and intentions to use mobile travel booking. 

People tend to share their travel experience on social media platforms during or after the trip~\cite{Fotis-J2012}, research has been conducted to learn the underline incentives and motivations. \citet{Myunghwa-K2013} argued that people are more likely to share if they are motivated to travel, satisfied with their trip, and trust social media, whereas \citet{Munar-A2014} found that people are motivated to share content for personal and community-related benefits and identified that the most popular content for sharing is visual content, including images and videos. They also found that sharing on social media is seen as a way to connect with others and provide emotional support rather than as a source of information for vacation planning. More specifically, \citet{Hubert-S2023} identified four factors, including blurred but clear views, positive incentives, subjective well-being, and restraining factors that can influence travellers' desire to share personal information with tourism websites and service providers. From a slightly different perspective, \citet{Grace-B2021} studied the impact of sharing their experiences during and after a trip on a traveller's overall life satisfaction and found that sharing experiences, such as posting pictures on social media, during a trip can potentially augment the positive impact of fulfilling holiday motivations on overall satisfaction. Similarly, sharing experiences, such as publishing blog posts, after a trip can amplify the positive influence of holiday trip satisfaction on subjective well-being.

Despite the usefulness and popularity of utilising such technologies to enhance the travel experience, \citet{Tussyadiah-I2019} argued such emerging technologies raise new layers of privacy concerns for tourists due to the lack of full transparency in how these technologies handle personal data, and highlighted that the adoption of new technologies in the tourism industry leads to increased risks to travellers' privacy. Regardless of the travel stage, people's attitudes and perceptions about privacy concerns and/or perceived risks related to personal data sharing in the context of tourism have frequently been the subject of extensive research. \citet{Ioannou-A2020} conducted studies to learn about travellers' perceptions of sharing their biometric and behavioural data with travel providers. The findings revealed that although travellers have concerns about their privacy, they are still open to sharing their behavioural data. However, their willingness to share biometric data depends on the expected benefits rather than privacy concerns. 

Such privacy-benefit/incentive trade-off has been further investigated in a study~\cite{Lu-Y2022} of utilising personal data flow modelling to evaluate privacy risks and concerns for travel booking. In addition, location data is also often regarded as sensitive data that can be used to infer behaviour patterns and other personal information. \citet{Ingolf-B2021} conducted research to examine travellers' attitudes towards location data sharing when using transportation services in the UK. They identified four different groups of people with varying attitudes towards location privacy. They suggested that these differences should be considered when designing location-based services and privacy-enhancing technologies.

Taking a different angle, there are other studies that focus on exploring factors that can influence people's attitudes and perceptions of privacy concerns. \citet{Francisco-F2022} suggested that factors such as the risk associated with using different types of technology, past experiences with data misuse, and lack of awareness of data management practices can have an impact on privacy concerns. In addition, \citet{Francisco-F2022} pointed out that travellers could adopt different strategies to protect their data in response to privacy concerns. To mitigate travellers' privacy concerns and perceived risks, \citet{Afolabi-2021} argued that the use-context of location-based service can help mitigate, suggesting that the tourist service providers should emphasise the use-context of technology to provide more information about how tourists' information will be used to gain more trust. In a more general context, \citet{Acquisti-A2015} highlighted the context-dependence of privacy preference, illustrating that individuals can vary from showing extreme concern to complete apathy regarding privacy issues in different situations. This observation has major implications for sharing travel experiences via social media, as people do not have a clear sense of the spatial boundaries to determine who reads or views content posted online. This adds complexity to privacy decision-making, especially when different social groups mix on the Internet.

Although the research literature has explored diverse aspects such as the use of technologies in different stages of leisure travel and travellers' attitudes and perceptions towards personal data sharing related privacy concerns, limited work has been conducted to understand how and to what extent different travel stages influence changes in leisure travellers' privacy attitudes towards sharing personal data. The purposes of sharing personal data vary significantly across different travel stages. Before travelling, data sharing may involve booking transportation or accommodations or sharing the news of upcoming holidays. During the trip, it could include booking local activities or updating social media with travel experiences. After travelling, it might involve providing feedback or reviews. These varying purposes can be considered as proxies for different travel stages, a perspective that has been under-explored.  Moreover, the possible impact of different travel stages on leisure travellers' behaviour of sharing travel experiences on different social media platforms is also overlooked. 

Having these in mind, to better answer \textbf{RQ1}, we derived two hypotheses:
\begin{itemize}
\item \textbf{H1}: \emph{Individual traveller's attitude towards sharing personal data differs depending on personal data sharing purposes.}

\item \textbf{H2}: \emph{Individual traveller's attitude towards sharing personal data differs depending on personal data types.} 
\end{itemize}

In addition, to better answer \textbf{RQ2}, we defined the following hypotheses:
\begin{itemize}
\item \textbf{H3}: \emph{Individual travellers exhibit different attitudes towards sharing travel-related content when using different social media platforms.}

\item \textbf{H4}: \emph{Individual travellers exhibit different attitudes towards sharing travel-related content at different stages of leisure travel.}
\end{itemize}

\section{Methodology}
\label{sec:methodology}

Since our RQs are around privacy-related attitudes and behaviours as perceived by leisure travellers, we decided to conduct an online survey to collect self-perceived data to address the RQs. In this section, the research methodology is presented, including the design of the survey (\ref{sec:design}), followed by introducing the data collection criteria (\ref{sec:data_collection}). In section~\ref{sec:data_analysis}, the methodology for the data analysis is described. Section~\ref{sec:ethics} presents the ethical considerations of this study. We adopted an exploratory approach for the data analysis, aiming to unearth insights that may have eluded previous studies within the field.

\subsection{Survey design}
\label{sec:design}

The online survey\footnote{The full survey can be accessed from \url{https://figshare.com/s/8f8116effd678b66e258}} is designed to have the following five sections with 40 questions in total, and three of these questions are used for validation purposes and are placed at the beginning, the middle, and the end of the survey, respectively. The full survey can be accessed from 

\subsubsection{Demographics} 
\label{sec:survey_p1}
This section includes questions to learn about travellers' gender, age group, marriage status, and ethnicity.

\subsubsection{General background}
\label{sec:survey_p2}

This section aims to understand travellers' general usage of some selected social media platforms (e.g., frequency of using social media platforms, sharing behaviour) and their usual travelling patterns. Based on the statistics reported in Wikipedia~\footnote{\url{https://en.wikipedia.org/wiki/List_of_social_platforms_with_at_least_100_million_active_users}}, we initially identified the top 20 social media platforms with over 100 million active users. Then we excluded social media platforms primarily used as instance messengers (IMs), such as WhatsApp, Meta Messenger, and Telegram. Additionally, considering the English fluency requirement for the online survey, we further excluded social media platforms predominantly used in China, such as WeChat, Kuaishou, and QQ. Furthermore, LinkedIn was excluded due to its focus on professional and career-oriented networking. Finally, our selection of social media platforms includes Facebook, Instagram, YouTube, Twitter (now known as X), TikTok, Snapchat, Pinterest. To introduce randomness and variability into the responses, we added Reddit, Flickr, and Spotify into the relevant questions. It is worth noting that these selected social media platforms were also used in questions in other sections of the survey.   
    
\subsubsection{Before leisure travel}
\label{sec:survey_p3}

This section covers questions about 
\begin{itemize}
    \item learning travellers' leisure travel booking practices in terms of how to choose a travel destination (e.g., internet search, visit travel agency, word of mouth, etc.), where to book a holiday (e.g., travel agency, online booking site, etc.), and who would normally book a holiday (e.g., traveller themselves, their family members, etc.)
    \item \label{sec:social_media_usage} learning travellers' usage of social media before leisure travel. Specifically, inspired by the study~\cite{Amaro-S2017}, questions about learning the purposes ((e.g., looking for inspiration, reading reviews, etc.) of using social media are included
    \item \label{sec:personal_data} learning how comfortable travellers feel towards personal data collection and sharing for holiday bookings. Following the recommendations in~\cite{nist800122} and reviewing the personal data types introduced in the study~\cite{Ioannou-A2021,Francisco-F2022}, we have incorporated 31 distinct types of personal data in this study (see Section~\ref{sec:privacy_attitudes} for more details). 
\end{itemize}

\subsubsection{During leisure travel}
\label{sec:survey_p4}

This section is designed to understand travellers' privacy attitudes toward sharing personal data for booking travel-related activities as well as how and to what extent travellers share their travel-related experiences on social media platforms during leisure travel. We developed questions based on the slight variation of the questions in Section~\ref{sec:survey_p3} item~\ref{sec:social_media_usage} and Section\ref{sec:survey_p3} item~\ref{sec:personal_data} so that we can examine the changes in travellers' attitudes and behaviour.

\subsubsection{After leisure travel}
\label{sec:survey_p5}

This section aims to capture travellers' behaviours of sharing their travel-related experience on social media platforms after leisure travel for summarising and reflecting on the travel. The questions introduced in Section~\ref{sec:survey_p3} item~\ref{sec:social_media_usage} and Section\ref{sec:survey_p3} item~\ref{sec:personal_data} with minor alterations are included in this section.

\subsection{Data collection}
\label{sec:data_collection}

The survey was developed and conducted using a local installation of the Community Edition of the online survey system LimeSurvey\footnote{\url{https://www.limesurvey.org/}}, and participants were recruited using the crowdsourcing platform Prolific\footnote{\url{https://www.prolific.co/}} in February 2023. The participants were at least 18 years old and fluent in English to be able to understand the survey questions and to give valuable responses. We did not apply any other exclusion criteria to capture a more representative sample of the general population, since we consider most adults to be leisure travellers from time to time. At the time of writing this paper, Prolific had a pool of 124,651 active users who met the inclusion criteria. According to the latest study~\cite{Douglas-B2023} that compares different online human participants recruitment sites, Prolific is considered as one of the top 2 sites delivering the highest quality data. 

\begin{table*}[!htbtb]
\centering
\begin{threeparttable}
\caption{Demographic results}
\label{tab:demographics_stats}
\begin{tabular}{cc||cc||cc||cc}
\hline
\textbf{Age Group} & \%  & \textbf{Education} & \% & \textbf{Employment Sector} & \% & \textbf{Ethnic Group*} & \%\\
\hline
18-24 & 28 & Less than high school & 1 & Private Sector & 52 & White\tnote{w} & 67\\
25-34 & 41 & High school & 29 & Government & 11 & Black\tnote{b} & 26\\
35-44 & 20 & Bachelor's degree or equivalent & 48 & Retired & 2 & Asian\tnote{a} & 2\\
45-54 & 7 & Master's degree or equivalent & 21 & Self employed & 13 & Arab & 0.6\\
55-64 & 3 & Doctoral degree or equivalent & 0.6 & Unemployed & 13 & Mixed\tnote{m}&  0.3\\
above 65 & 1 & Prefer not to say & 0.4 & Other & 8 & Other & 0.1\\
\hline
\end{tabular}
\begin{tablenotes}
\small
\item[*] The categorisation is adopted from \url{https://www.ethnicity-facts-figures.service.gov.uk/style-guide/ethnic-groups}; [w] White (English, Welsh, Scottish, Northern Irish or British, Irish, Traveller, Roma, Any other White background)
; [b] Black (Caribbean, African, Any other Black and Caribbean background))
; [a] Asian (Indian, Pakistani, Bangladeshi, Chinese, Any other Asian background)
; [m] Mixed or multiple ethnic groups (White and Black Caribbean, White and Black African, White Asian, Any other Mixed or multiple ethnic backgrounds)
\end{tablenotes}
\end{threeparttable}
\end{table*}

In this study, each participant received financial compensation at the rate of £9/hour for their time spent on taking the survey, which took an average of 22 minutes to complete. To best estimate the number of participants needed for this study, a power analysis using G*Power~\cite{Faul-F2007} was carried out. For testing the statistical differences between two groups of data with equal size (allocation ratio = 1), the recommendation is to have a sample size of at least 282 participants for detecting a medium effect size ($f = 0.3$) at 80\% power ($\alpha = 0.05$). For the analysis of variance (ANOVA) for three groups of data (i.e., `before leisure travel', `during leisure travel', and `after leisure travel'), the power analysis suggested that a sample size of at least 159 participants is needed to detect a medium effect size ($f = 0.25$) at 80\% power ($\alpha = 0.05$). To be on the side of caution, we decided to recruit 330 participants.

\subsection{Data analysis}
\label{sec:data_analysis}

A total of 322 individuals worldwide were recruited to participate in this study. Four participants were excluded as they failed to answer the validation questions, leaving responses from 318 participants for subsequent data analysis. In addition, all responses are converted from text to numerical numbers, and responses with the rating of ``no response'' or ``no value'' were excluded as missing values from the later analysis. Mixed methods were used for the data analysis using Microsoft Excel and Python module SciPy\footnote{\url{https://scipy.org/}} and statsmodels\footnote{\url{https://www.statsmodels.org/}}. We adopted the exploratory data analysis approach to examine the distribution of responses to different questions from the ``General background'' section of the survey, aiming to learn the demographics and general usage of social media platforms, providing some background information for further analysis and comparison. To answer the research questions and test the four outlined hypotheses, we began by testing the data normality using the Shapiro-Wilk test. Upon confirming that the collected data did not pass the normality test, we employed the Mann-Whitney U tests to examine statistical differences between two independent groups, the Wilcoxon signed-rank tests to compare statistical differences between two paired groups, and Chi-square tests to compare the distribution of aggregated data from two groups.

In addition, the Aligned Rank Transform (ART) methodology was applied to transform the non-parametric data, so it can be used by a three-way ANOVA with Type ||| sum of squares (SS) to examine the effect of travel stages on participants' self-reported privacy attitudes. Moreover, a multivariate logistic regression model was applied to investigate the determining factors that can affect how our survey participants reported their behaviours on sharing travel-related content on social media platforms across different travel stages. Furthermore, different data visualisation techniques such as the heat map and stacked bar chart were utilised to help us discover hidden insights and patterns.

\subsection{Ethics consideration}
\label{sec:ethics}

This study received a favourable ethical opinion from the Central Research Ethics Advisory Group of the University of Kent (Reference Number: CREAG013-11-22) and was conducted following the requirements of the ethics committee. These requirements contain a range of considerations, including but not limited to, avoiding inducing psychological stress or harm, as well as refraining from directly or indirectly identifying individuals. We collected demographic data including age group, gender, education, employment, and ethnic group in this study. No personal identifiable data such as name/email address were collected. All data were stored on a secure server hosted at the relevant institution and only the members of the research team were given access. All participants were transparently informed about the study's objectives and procedure through a participant information sheet and subsequently consented electronically before proceeding to take the online survey.

\section{Results}
\label{sec:analysis_results}

This section first presents descriptive results of exploratory data analysis to learn the demographics of the recruited participants and establish an understanding of travellers' general usage of social media platforms (Section~\ref{sec:decriptive_results}). Then, Sections~\ref{sec:privacy_attitudes} and \ref{sec:sharing} present results to answer RQ1 and RQ2, respectively.

\subsection{Descriptive results}
\label{sec:decriptive_results}

\begin{table*}[!tb]
\caption{A heat-map showing the participants' self-reported travelling patterns with different travelling companionship types}
\label{table:travel_experience}
\centering
\pgfplotstabletypeset[%
     color cells={min=0,max=200,textcolor=black},
     /pgfplots/colormap={}{rgb255=(255,255,255) rgb255=(255,0,0)},
    /pgf/number format/fixed,
    /pgf/number format/precision=3,
    col sep=comma,
    columns/-/.style={reset styles,string type}%
]{
-, Alone, Colleagues, Friends, Family without kids, Family with kids
Once per week, 20, 6, 10, 9, 14
Once per month, 48, 20, 37, 26, 20
Once per quarter, 33, 33, 64, 42, 47
Once per half year, 25, 28, 62, 49, 54
Once per year, 43, 47, 67, 62, 65
Less than once per year, 114, 122, 58, 94, 63
}
\end{table*}

\subsubsection{Demographics}

A total of 322 individuals worldwide were recruited to participate in this study. After the validation check, 4 participants were excluded and the responses from 318 individuals (167 females, 149 males, and 2 preferred not to say) were included in the data analysis. As depicted in Table~\ref{tab:demographics_stats}, 41\% of participants were aged between 25-34, and nearly half (48\%) of the participants reported having a Bachelor's degree or equivalent. Just over half of the participants were working in the private sector, followed by 13\% of participants who were either self-employed or unemployed. The majority (67\%) of the participants reported to be White, and nearly one-third (26\%) of the population are Black. By looking at the countries of residence for all participants, two-thirds (66.4\%) of the participants are from Europe, nearly one-third (30.1\%) are from Africa, and the remaining participants (3.5\%) are from Asia, Oceania and North America. Table~\ref{tab:country_residence} provides further details, indicating that European participants reside in over 15 countries, while all African participants are from South Africa.

\begin{table*}[!htb]
\caption{Number of participants per country}
\label{tab:country_residence}
\centering
\begin{tabular}{c c|c c|c c|c c}
\toprule
South Africa & 97 & Greece & 14 & Australia & 2 & Finland & 1\\
Poland & 54 & Spain & 10 & Czech Republic & 2 & Sweden & 1\\
Portugal & 46 & Estonia & 5 & Belgium & 2 & Slovenia & 1\\
United Kingdom & 25 & Israel & 5 & Denmark & 2 & Mexico & 1\\
Hungary & 20 & France & 5 & New Zealand & 1 & Germany & 1\\
Italy & 18 & Netherlands & 3 & Latvia & 1 & Unknown & 1\\
\bottomrule
\end{tabular}
\end{table*}

\begin{table}[!htb]
\centering
\begin{threeparttable}
\caption{U statistics and $p$-values of multiple Mann-Whitney U tests for comparing travelling patterns between female and male participants, and between European and South African participants}
\label{tab:pair_travel_experience}
\begin{tabular}{c c c}
\toprule
Companionship types & F vs.\ M S(p) & EU vs.\ SA S(p)\\
\midrule
Alone & 13280 (0.132) & 12500 (*)\\
With colleagues & 13601 (0.017) & 11492 (0.012)\\
With friends & 12372 (0.817) & 12726 (*)\\
With family without kids & 12111 (0.922) & 10878 (0.175)\\
With family with kids & 11890 (0.904) & 11593 (0.009)\\
\bottomrule
\end{tabular}
\begin{tablenotes}
\small
\item Notes: F is for female travellers; M is for male travellers; S(p) is for U Statistic ($p$-value); EU stands for Europe; SA stands for South Africa; * = $p \le 0.001$.\\
\end{tablenotes}
\end{threeparttable}
\end{table}

\subsubsection{General travel behaviours}

To understand participants' general travel behaviour, we asked participants about their travelling frequency with different travelling companionship types (i.e., travel alone, travel with colleagues, travel with friends, travel with family with kids, and travel with family without kids). As demonstrated in Table~\ref{table:travel_experience}, participants reported varying frequencies of solo travel, with the highest frequency being ``Less than once per year'' (i.e., 114) and the lowest being ``Once per week'' (i.e., 20). Participants frequently travelled with friends, with the highest frequency being ``Once per year'' (i.e., 67) and the lowest being ``Once per week'' (i.e., 10). In addition, participants exhibited similar patterns when travelling with family members with kids or without kids. Compared to other companionship types, travelling with colleagues was less common. with the highest frequency reported as ``Less than once per year'' (i.e., 122) and the lowest as ``Once per week'' (i.e., 6). Overall, participants exhibited diverse travel patterns across different companionship types, with the majority opting for less frequent travel with colleagues and more frequent travel with friends and family, both with and without kids. In addition, other factors such as gender, social platform, countries/culture~\cite{Karatsoli-M2020, Ioannou-A2020, Afolabi-2021} have been reported to be associated with different privacy attitudes and data sharing decision-making in the tourism domain. 
Here in this study, we also looked at the self-reported general travel behaviour from these perspectives. The statistical test results reported in Table~\ref{tab:pair_travel_experience} suggest that there are no significant differences in travelling frequencies between female and male participants across all travel companionship types (see Table~\ref{tab:male_travel_experience} and Table~\ref{tab:female_travel_experience} in Section~\ref{sec:appendix} for the visualisation of using heat-maps to show travelling patterns with different travelling companionship types for male and female travellers respectively). 

Furthermore, we compared such travel patterns of participants from Europe and those from South Africa, and slightly different observations were obtained. As reported in Table~\ref{tab:pair_travel_experience}, participants from Europe and those from South Africa exhibited statistically different travelling patterns when travelling alone and travelling with friends (visualised in the heat maps in Tables~\ref{table:travel_experience_eu} and \ref{table:travel_experience_sa}), suggesting the importance of considering geographic differences in further analysis.

Given that each participant had the option to select multiple choices to represent their travelling patterns, we considered the most frequently reported frequency as the primary indicator of each participant's travelling patterns, irrespective of the travel companionship type. This approach allowed us to establish each participant's previous travel frequency profile. Specifically, individuals reporting their highest frequency as ``Once per week'', ``Once per month'', or ``Once per quarter'' were categorised as frequent travellers, while those selecting other options were classified as less frequent travellers. We were interested to see how such profiles would help us understand and analyse travellers' responses to the survey questions along with the gender and country of residence variables for the remaining analyses.

\subsubsection{General usage of social media platforms}
\label{sec:general_use}

To understand the general use of selected social media platforms, we asked participants ``\emph{How often do you use the social media platforms listed below?}''. By closely looking at the overall responses across all participants as shown in Table~\ref{tab:general_usage_social}, Facebook, Instagram and YouTube are the popular social media platforms that participants frequently visit multiple times per day, followed by Twitter and TikTok. Whereas, Pinterest and Snapchat are the less popular ones among all participants. Then we compared the participants' responses between 1) the frequent traveller group with the less frequent traveller group; 2) the male travellers and female travellers; and 3) the European travellers and South African travellers, for each selected social media platform. 

\begin{table*}[!htb]
\caption{A heat-map showing the self-reported usage of different social media platforms}
\label{tab:general_usage_social}
\centering
\pgfplotstabletypeset[%
     color cells={min=0,max=300,textcolor=black},
     /pgfplots/colormap={}{rgb255=(255,255,255) rgb255=(255,0,0)},
    /pgf/number format/fixed,
    /pgf/number format/precision=3,
    col sep=comma,
    columns/-/.style={reset styles,string type}%
]{
-, Facebook, Instagram, Twitter, TikTok, Snapchat, YouTube, Pinterest
Multiple times per day, 145, 148, 102, 87, 14, 184, 27
At least once per day, 53, 52, 30, 25, 12, 56, 23
Multiple times per week, 33, 33, 28, 28, 16, 41, 30
Once per week, 24, 20, 17, 14, 8, 25, 25
Less frequent than once per week, 47, 32, 76, 54, 101, 11, 112
Never used before, 16, 33, 65, 110, 165, 0, 101
}
\end{table*}

\begin{table}[!htb]
\centering
\begin{threeparttable}
\caption{U statistics and $p$-values of multiple pairwise Mann-Whitney U tests for comparing usage of social media platforms}
\label{tab:usage_social_travel_comparison}
\small
\begin{tabular}{c c c c}
\toprule
Social Media & FT vs.\ L-FT S(p) & F vs.\ M S(p) & EU vs.\ SA S(p)\\
\midrule
Facebook & 8920 (0.004) & 12482	(0.958) & 11231 (0.147)\\
Instagram & 10668 (0.664) & 9676 (*) & 12557 (*)\\
Twitter & 8163 (*) & 12120 (0.684) & 15212 (*)\\
TikTok & 9785 (0.100) & 9018 (*) & 15101 (*)\\
Snapchat & 8950	(0.005) & 9444 (*) & 13202 (*)\\
YouTube & 10336	(0.412) & 15637	(*) & 12687 (*)\\
Pinterest & 10159 (0.256) & 7596 (*) & 13076 (*)\\
\bottomrule
\end{tabular}
\begin{tablenotes}
\small
\item FT is for frequent travellers; L-FT is for less-frequent travellers; 
\\F is for female travellers; M is for male travellers; 
\\EU is for Europe; 
\\SA is for South Africa; 
\\S(p) is for U-Statistic ($p$-value), * = $p \le 0.001$\\
\end{tablenotes}
\end{threeparttable}
\end{table}

As shown in Table~\ref{tab:usage_social_travel_comparison}, aside from differences in Twitter usage, we found no significant distinctions between frequent and less frequent participants for other social media platforms. As further illustrated in Table~\ref{tab:usage_social_travel_experience}, frequent travellers use Twitter more often compared to less frequent travellers. Differently, notable differences were observed between female and male participants. Female and male participants exhibited significantly different behaviours across all social media platforms except Twitter and Facebook. As further indicated in Table~\ref{tab:usage_social_gender}, male participants exhibited a higher frequency of using Instagram, TikTok, and Pinterest than female participants, whereas female participants tended to visit YouTube more frequently than male participants. In addition, participants from Europe and South Africa displayed different behaviours using most of the selected social media platforms, apart from Facebook. As further demonstrated in Table~\ref{tab:usage_social_country_residence}, participants from South Africa reported more engagement in using social media platforms than participants from Europe in general.

In summary, we found that people's previous travel frequency (i.e., frequent travellers vs.\ less frequent travellers) would not affect how often people use the selected social media platforms in general, whereas gender and country of residency do influence usage patterns on most selected platforms. We would like to highlight that our work did not focus on investigating the underlying causes of these observations, since such an investigation would require a very different methodology. However, the results from such exploratory data analyses offered insights that can inform our further analyses, where gender, country of residence, and previous travel frequency were taken into consideration while addressing RQ1 and RQ2.

\begin{table*}[!htb]
\caption{Self-reported usage of different social media platforms between frequent travellers (FT) and less frequent travellers (L-FT) in percentage}
\label{tab:usage_social_travel_experience}
\centering
\small
\begin{tabular}{l*{14}{c}}
\toprule
& \multicolumn{2}{c}{Facebook} & \multicolumn{2}{c}{Instagram} & \multicolumn{2}{c}{Twitter} & \multicolumn{2}{c}{TikTok} & \multicolumn{2}{c}{Snapchat} & \multicolumn{2}{c}{YouTube} & \multicolumn{2}{c}{Pinterest} \\
\cmidrule(lr){2-3} \cmidrule(lr){4-5} \cmidrule(lr){6-7} \cmidrule(lr){8-9} \cmidrule(lr){10-11} \cmidrule(lr){12-13} \cmidrule(lr){14-15}
- & \small{FT} & \small{L-FT} & \small{FT} & \small{L-FT} & \small{FT} & \small{L-FT} & \small{FT} & \small{L-FT} & \small{FT} & \small{L-FT} & \small{FT} & \small{L-FT} & \small{FT} & \small{L-FT} \\
\midrule
Multiple times per day&0.53&0.36&0.50&0.43&0.39&0.20&0.31&0.25&0.05&0.03&0.60&0.55&0.11&0.05\\
At least once per day&0.16&0.18&0.15&0.20&0.13&0.05&0.09&0.08&0.05&0.03&0.16&0.19&0.09&0.04\\
Multiple times per week&0.08&0.14&0.08&0.14&0.07&0.10&0.10&0.08&0.07&0.03&0.14&0.13&0.09&0.10\\
Once per week&0.07&0.08&0.08&0.04&0.05&0.06&0.04&0.06&0.03&0.02&0.08&0.09&0.06&0.10\\
Less than once per week&0.13&0.19&0.09&0.12&0.18&0.34&0.18&0.15&0.34&0.29&0.03&0.04&0.34&0.40\\
Never used before&0.04&0.06&0.11&0.07&0.18&0.25&0.30&0.39&0.46&0.61&0.00&0.00&0.30&0.31\\
\bottomrule
\end{tabular}
\end{table*}

\begin{table*}[!htb]
\caption{Self-reported usage of different social media platforms between male (M) and female (F) travellers in percentage}
\label{tab:usage_social_gender}
\centering
\small
\begin{tabular}{l*{14}{c}}
\toprule
& \multicolumn{2}{c}{Facebook} & \multicolumn{2}{c}{Instagram} & \multicolumn{2}{c}{Twitter} & \multicolumn{2}{c}{TikTok} & \multicolumn{2}{c}{Snapchat} & \multicolumn{2}{c}{YouTube} & \multicolumn{2}{c}{Pinterest} \\
\cmidrule(lr){2-3} \cmidrule(lr){4-5} \cmidrule(lr){6-7} \cmidrule(lr){8-9} \cmidrule(lr){10-11} \cmidrule(lr){12-13} \cmidrule(lr){14-15}
- & M & F & M & F & M & F & M & F & M & F & M & F & M & F \\
\midrule
Multiple times per day&0.47&0.44&0.58&0.34&0.36&0.28&0.39&0.15&0.05&0.04&0.46&0.70&0.11&0.05\\
At least once per day&0.16&0.17&0.11&0.23&0.08&0.11&0.09&0.07&0.04&0.03&0.22&0.13&0.13&0.01\\
Multiple times per week&0.08&0.13&0.10&0.11&0.05&0.13&0.07&0.11&0.07&0.03&0.15&0.11&0.12&0.07\\
Once per week&0.07&0.08&0.03&0.10&0.08&0.03&0.03&0.06&0.04&0.01&0.11&0.05&0.08&0.07\\
Less than once per week&0.17&0.12&0.10&0.10&0.21&0.26&0.14&0.20&0.40&0.23&0.06&0.01&0.39&0.31\\
Never used before&0.05&0.05&0.08&0.13&0.22&0.19&0.28&0.42&0.41&0.64&0.00&0.00&0.17&0.48\\
\bottomrule
\end{tabular}
\end{table*}

\begin{table*}[!htb]
\caption{Self-reported usage of different social media platforms between Europe (EU) and South Africa (SA) travellers in percentage}
\centering
\small
\begin{tabular}{l*{14}{c}}
\toprule
& \multicolumn{2}{c}{Facebook} & \multicolumn{2}{c}{Instagram} & \multicolumn{2}{c}{Twitter} & \multicolumn{2}{c}{TikTok} & \multicolumn{2}{c}{Snapchat} & \multicolumn{2}{c}{YouTube} & \multicolumn{2}{c}{Pinterest} \\
\cmidrule(lr){2-3} \cmidrule(lr){4-5} \cmidrule(lr){6-7} \cmidrule(lr){8-9} \cmidrule(lr){10-11} \cmidrule(lr){12-13} \cmidrule(lr){14-15}
- & EU & SA & EU & SA & EU & SA & EU & SA & EU & SA & EU & SA & EU & SA \\
\midrule
Multiple times per day&0.42&0.53&0.42&0.61&0.20&0.60&0.18&0.49&0.05&0.03&0.52&0.72&0.05&0.15\\
At least once per day&0.16&0.16&0.17&0.15&0.09&0.09&0.07&0.10&0.03&0.05&0.18&0.19&0.05&0.12\\
Multiple times per week&0.12&0.06&0.11&0.07&0.09&0.06&0.09&0.09&0.02&0.12&0.16&0.05&0.08&0.13\\
Once per week&0.09&0.06&0.07&0.04&0.04&0.08&0.05&0.04&0.00&0.07&0.09&0.04&0.09&0.05\\
Less frequent than once per week&0.17&0.10&0.10&0.08&0.30&0.09&0.15&0.18&0.27&0.39&0.05&0.00&0.36&0.34\\
Never used before&0.04&0.08&0.13&0.04&0.27&0.07&0.46&0.09&0.62&0.31&0.00&0.00&0.36&0.20\\

\bottomrule
\end{tabular}
\label{tab:usage_social_country_residence}
\end{table*}

\subsection{Privacy attitudes towards the sharing of different types of personal data for different intended purposes}
\label{sec:privacy_attitudes}

To answer \textbf{RQ1}, we asked participants three questions about their attitudes toward sharing various personal data types (PDTs) for different personal data sharing purposes (PURPOSEs) during different phases of leisure travel. The first question is:

``\emph{The list below includes different types of personal information. For each item, please tell us how comfortable you are for the travel companies to collect this information.}''

It aimed to gain insights into a traveller's stance regarding sharing personal data with online booking sites during the process of making leisure travel bookings. This is considered the first personal data sharing purpose (PURPOSE 1). The second question is: 

``\emph{The list below includes different types of personal information. For each item, please tell us how comfortable you are for the travel companies to share this information with other third parties/services (e.g., for renting a car, booking a table, insurance, etc.)}''

This question is designed to explore travellers' attitudes on sharing personal data with other entities, including subsidiaries and associated third-party services, affiliated with the online booking site that they use for their leisure travel bookings. The main purpose is to use these affiliated entities to facilitate the online booking process to provide extra benefits to travellers. We consider this as PURPOSE 2. The third question is: 

``\emph{The list below includes different types of personal information. For each item, please tell us how comfortable you are sharing your information to enrich your travel experience during your holiday (e.g.., get the information and/or booking services that you need, searching for information and/or booking trip-related services such as booking tickets for the theatre, or booking a table in a restaurant).}''

This question is about PURPOSE 3, which aims to learn travellers' attitudes towards sharing personal data for booking various activities and services that could enhance the overall travel experience during their leisure trips. These activities may include making reservations for dining at a restaurant or purchasing tickets for theatrical performances, among others. For all the above questions, participants' attitudes are measured using 7-point Likert scale regarding the level of comfortableness in sharing personal data. For convenience, we use the comfortableness score (C) for the rest of this paper. A score of 1 indicates that the traveller is very uncomfortable sharing personal data, whereas a score of 7 indicates that the traveller feels very comfortable sharing personal data.

Inspired by the recommendations in~\cite{nist800122, Ioannou-A2021, Francisco-F2022}, we included 31 types of personal data in this survey, including name (NAME), gender (GEND), date of birth (DOB), nationality (NATI), home address (HOME), email address (EMAIL), phone number (PHONE), height (HEIG), weight (WEIG), sexual orientation (SEXO), Ethnicity (ETHN), religion (RELIG), education (EDUC), profession (PROF), credit card information (CRED), bank account information (BANK), contacts (CONT), passport information (PASS), driver license information (DRIV), vehicle registration number (VRN), medical information (MEDIC), face image (FACE), voice sample (VOICE), social media profile data (SOCIA), Internet Protocol (IP) address (IP), Media Access Control (MAC) address (MAC), Web-usage data (i.e., web browsing records) (WEB), real time location data (LOC), activity sensor data (ACTIV), personal interests (PERSO), and dietary requirements (DIET).

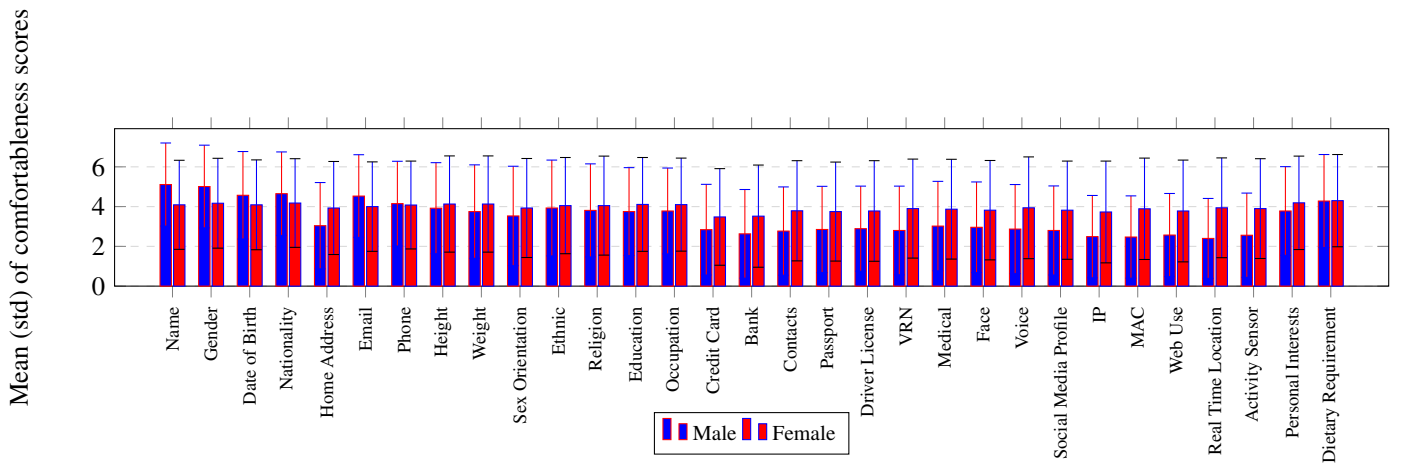
\begin{figure*}[!htb]
\caption{Bar chart showing mean and standard deviation of comfortableness scores of sharing different data types for PURPOSE 3 from both male and female participants}
\label{fig:female_male_in_holiday_mean}
\centering
\begin{tikzpicture}
\begin{axis}[
    width=\linewidth,
    height=0.2\linewidth,
    ylabel={Mean (std) of comfortableness scores},
    bar width=0.15cm,
    ybar=2*\pgflinewidth,
    ymajorgrids = true,
    grid style = {dashed,gray!30},
    legend style={at={(0.5,-0.8)},anchor=north,legend columns=-1,
    },
    error bars/y dir=both,
    error bars/y explicit,
    xtick=data,
    xticklabels={Name, Gender, Date of Birth, Nationality, Home Address, Email, Phone, Height, Weight,
            Sex Orientation, Ethnic, Religion, Education, Occupation, Credit Card, Bank,
            Contacts, Passport, Driver License, VRN, Medical, Face, Voice,
            Social Media Profile, IP, MAC, Web Use, Real Time Location, Activity Sensor,
            Personal Interests, Dietary Requirement},
    x tick label style={rotate=90, anchor=east, font=\fontsize{7}{8}\selectfont},
    enlarge x limits=0.05,
    ymin=0,
    legend cell align=left,
    legend style={font=\footnotesize},
]
\addplot[
    ybar,
    bar width=0.15cm,
    color=blue,
    fill=blue,draw=red,
    error bars/.cd,
    y explicit
] 
coordinates {
    (0,5.11) +- (2.09,2.09)
    (1,5.01) +- (2.08,2.08)
    (2,4.57) +- (2.20,2.20)
    (3,4.65) +- (2.10,2.10)
    (4,3.04) +- (2.17,2.17)
    (5,4.53) +- (2.08,2.08)
    (6,4.15) +- (2.13,2.13)
    (7,3.92) +- (2.29,2.29)
    (8,3.75) +- (2.35,2.35)
    (9,3.53) +- (2.50,2.50)
    (10,3.93) +- (2.41,2.41)
    (11,3.81) +- (2.34,2.34)
    (12,3.75) +- (2.21,2.21)
    (13,3.78) +- (2.16,2.16)
    (14,2.84) +- (2.28,2.28)
    (15,2.63) +- (2.23,2.23)
    (16,2.77) +- (2.22,2.22)
    (17,2.85) +- (2.17,2.17)
    (18,2.89) +- (2.14,2.14)
    (19,2.80) +- (2.23,2.23)
    (20,3.02) +- (2.25,2.25)
    (21,2.96) +- (2.28,2.28)
    (22,2.87) +- (2.24,2.24)
    (23,2.80) +- (2.24,2.24)
    (24,2.49) +- (2.07,2.07)
    (25,2.47) +- (2.07,2.07)
    (26,2.57) +- (2.09,2.09)
    (27,2.40) +- (2.01,2.01)
    (28,2.56) +- (2.12,2.12)
    (29,3.78) +- (2.23,2.23)
    (30,4.28) +- (2.34,2.34)
};
\addlegendentry{Male}
\addplot[
    ybar,
    bar width=0.15cm,
    fill=red, draw=blue,
    error bars/.cd,
    y explicit
] 
coordinates {
    (0,4.09) +- (2.24,2.24)
    (1,4.17) +- (2.26,2.26)
    (2,4.09) +- (2.26,2.26)
    (3,4.18) +- (2.23,2.23)
    (4,3.93) +- (2.34,2.34)
    (5,4.00) +- (2.25,2.25)
    (6,4.08) +- (2.21,2.21)
    (7,4.13) +- (2.42,2.42)
    (8,4.13) +- (2.42,2.42)
    (9,3.93) +- (2.49,2.49)
    (10,4.05) +- (2.42,2.42)
    (11,4.05) +- (2.49,2.49)
    (12,4.11) +- (2.36,2.36)
    (13,4.10) +- (2.34,2.34)
    (14,3.48) +- (2.43,2.43)
    (15,3.52) +- (2.57,2.57)
    (16,3.79) +- (2.52,2.52)
    (17,3.75) +- (2.49,2.49)
    (18,3.78) +- (2.53,2.53)
    (19,3.90) +- (2.49,2.49)
    (20,3.87) +- (2.51,2.51)
    (21,3.82) +- (2.50,2.50)
    (22,3.94) +- (2.56,2.56)
    (23,3.82) +- (2.47,2.47)
    (24,3.73) +- (2.56,2.56)
    (25,3.89) +- (2.55,2.55)
    (26,3.78) +- (2.56,2.56)
    (27,3.94) +- (2.51,2.51)
    (28,3.90) +- (2.51,2.51)
    (29,4.19) +- (2.35,2.35)
    (30,4.30) +- (2.32,2.32)
};
\addlegendentry{Female}
\end{axis}
\end{tikzpicture}
\end{figure*}

\begin{table}[!htb]
\centering
\caption{U statistic results of multiple pairwise Mann-Whitney U tests to compare self-reported attitudes toward personal data sharing between female and male travellers for different personal data sharing purposes}
\label{tab:pair_female_male_personal_data_sharing}
\small 
\begin{tabular}{c c c c}
\toprule
PDT & \multicolumn{3}{c}{U Statistic (p-value)}\\
\midrule
& PURPOSE 1 & PURPOSE 2 & PURPOSE 3\\
\midrule
NAME & 12823 (0.466) & 11951.5 (0.830) & 15041 (*)\\
GEND & 12068.5 (0.781) & 11182.5 (0.228) & 14416 (0.002)\\
DOB & 12376 (0.896) & 11386 (0.348) & 13252 (0.117)\\
NATI & 12210.5 (0.934) & 10230.5 (0.015) & 13299 (0.103)\\
HOME & 11252 (0.197) & 10528.5 (0.046) & 8855.5 (*)\\
EMAIL & 11435.5 (0.322) & 11487 (0.418) & 13103.5 (0.110)\\
PHONE & 11197.5 (0.202) & 10640 (0.071) & 11915.5 (0.879)\\
HEIG & 9860.5 (0.002) & 10303 (0.019) & 11065 (0.211)\\
WEIG & 9328 (*) & 10055.5 (0.007) & 10553 (0.055)\\
SEXO & 10355.5 (0.018) & 10407.5 (0.035) & 10520.5 (0.059)\\
ETHN & 10483 (0.023) & 11044.5 (0.271) & 11314 (0.400)\\
RELIG & 10949.5 (0.112) & 10779.5 (0.097) & 10798.5 (0.134)\\
EDUC & 11081.5 (0.155) & 11249.5 (0.349) & 10588 (0.076)\\
PROF & 11252.5 (0.196) & 11067 (0.273) & 10642 (0.089)\\
CRED & 13188.5 (0.201) & 11829.5 (0.684) & 9602.5 (0.002)\\
BANK & 12121 (0.842) & 11685 (0.604) & 9167 (*)\\
CONT & 11554.5 (0.354) & 11595.5 (0.466) & 8580.5 (*)\\
PASS & 12361 (0.913) & 11585 (0.605) & 9109.5 (*)\\
DRIV & 11888 (0.691) & 11379.5 (0.371) & 8963.5 (*)\\
VRN & 11099 (0.133) & 10714.5 (0.054) & 8539 (*)\\
MEDIC & 11338.5 (0.265) & 11450.5 (0.360) & 9155 (*)\\
FACE & 10285.5 (0.010) & 10176.5 (0.011) & 9156 (*)\\
VOICE & 11415.5 (0.256) & 10899.5 (0.147) & 8679 (*)\\
SOCIA & 11126.5 (0.133) & 11271.5 (0.273) & 8694.5 (*)\\
IP & 11140 (0.131) & 11132 (0.211) & 7931 (*)\\
MAC & 11643 (0.402) & 11114 (0.173) & 7833.5 (*)\\
WEB & 12667 (0.606) & 11704.5 (0.623) & 8407 (*)\\
LOC & 11625.5 (0.589) & 10939 (0.086) & 7309 (*)\\
ACTIV & 11290 (0.359) & 10816.5 (0.089) & 7675.5 (*)\\
PERSO & 11828 (0.573) & 11214.5 (0.234) & 10293 (0.026)\\
DIET & 13379 (0.106) & 10835.5 (0.097) & 11660 (0.696)\\
\bottomrule
\end{tabular}
\begin{tablenotes}
\small
\item Notes: * = $p \le 0.001$.
\end{tablenotes}
\end{table}

\subsubsection{Previous travel frequency, gender, and country of residence}

First, we were keen to know if previous travel frequency plays a role in distinguishing travellers' attitudes toward personal data sharing, we applied Mann-Whitney U tests to compare collective responses for each PDT between frequent travellers and less-frequent travellers for each PURPOSE. 

First, we looked into the differences between female and male travellers (see Mann-Whitney U test results in Table~\ref{tab:pair_female_male_personal_data_sharing}). For PURPOSE 1 (i.e., data sharing during the booking process) and PURPOSE 2 (i.e., data sharing with affiliated entities during the booking process), there are no significant differences between female and male participants' attitudes toward sharing different data types. However, for PURPOSE 3 (i.e., data sharing for booking other activities during leisure travel), female and male participants exhibited significantly different attitudes towards half of the listed data types. To better visualise the differences, we plotted the means and standard deviations of responses from both female and male participants for PURPOSE 3 in Figure~\ref{fig:female_male_in_holiday_mean}, where female participants are more relaxed about sharing some personal data (i.e., home address (HOME), bank account information (BANK), contacts (CONT), passport information (PASS), driver license information (DRIV), vehicle registration number (VRN), medical information (MEDIC), face image (FACE), voice sample (VOICE), social media profile data (SOCIA), Internet Protocol (IP) address (IP), Media Access Control (MAC) address (MAC), Web-usage data (i.e., web browsing records) (WEB), real time location data (e.g., GPS data) (LOC), activity sensor data (ACTIV)) for booking other activities during a leisure travel. The contrast observed here suggests the potential gender-based differences in risk perception and trust, which may be shaped by varying levels of concern regarding privacy during travel.

In addition, we did the same analysis between frequent travellers and less frequent travellers, and between European and South African travellers. As shown in Table~\ref{tab:pair_frequent_less_frequent_personal_data_sharing}, previous travel frequency does not have an impact on how people perceive personal data sharing for different purposes. As depicted in Table~\ref{tab:pair_eu_sa_personal_data_sharing}, participants from Europe and those from South Africa largely had the same attitudes towards data sharing for most of the listed data types. The only two exceptions are contacts (CONT) and dietary requirements (DIET).

\begin{table}[!htbtb]
\centering
\caption{U statistic results of multiple pairwise Mann-Whitney U tests to compare self-reported attitudes toward personal data sharing between frequent travellers and less-frequent travellers for different personal data sharing purposes}
\label{tab:pair_frequent_less_frequent_personal_data_sharing}
\small 
\begin{tabular}{c c c c}
\toprule
PDT & \multicolumn{3}{c}{U Statistic (p-value)}\\
\midrule
& PURPOSE 1 & PURPOSE 2 & PURPOSE 3\\
\midrule
NAME&12679.5 (0.011)&12720 (0.007)&10762.5 (0.871) \\
GEND&12373 (0.033)&12657.5 (0.01)&10378 (0.705) \\
DOB&12065 (0.107)&11982.5 (0.101)&9694.5 (0.184) \\
NATI&12147.5 (0.081)&12385 (0.028)&10281 (0.609) \\
HOME&10524.5 (0.595)&10595 (0.839)&9732 (0.201) \\
EMAIL&13072.5 (0.002)&13056.5 (0.002)&10932.5 (0.542) \\
PHONE&12255 (0.067)&12040.5 (0.062)&10025 (0.389) \\
HEIG&10324 (0.42)&10859.5 (0.931)&9243 (0.05) \\
WEIG&10438 (0.566)&10704.5 (0.897)&9164.5 (0.037) \\
SEXO&10738 (0.87)&11252 (0.379)&9274.5 (0.062) \\
ETHN&11245.5 (0.651)&11368 (0.247)&9837 (0.29) \\
RELIG&10988.5 (0.819)&11579 (0.27)&9838 (0.291) \\
EDUC&10462.5 (0.59)&10573 (0.881)&9307.5 (0.08) \\
PROF&10389.5 (0.474)&10812 (0.851)&9234 (0.049) \\
CRED&10037.5 (0.277)&10709.5 (0.895)&9277.5 (0.061) \\
BANK&10039 (0.218)&10453.5 (0.66)&9793 (0.251) \\
CONT&11429.5 (0.474)&11154.5 (0.591)&9932 (0.373) \\
PASS&11815 (0.22)&11009 (0.627)&9831 (0.342) \\
DRIV&11081 (0.758)&10898.5 (0.817)&9879 (0.28) \\
VRN&10787.5 (0.861)&10812 (0.983)&9869 (0.306) \\
MEDIC&11179.5 (0.652)&11036 (0.725)&10043.5 (0.395) \\
FACE&12152 (0.085)&11037.5 (0.552)&10186.5 (0.57) \\
VOICE&11583.5 (0.34)&10907.5 (0.618)&10398 (0.724) \\
SOCIA&11952.5 (0.141)&11889.5 (0.07)&10155.5 (0.539) \\
IP&11360.5 (0.522)&10430.5 (0.729)&9494.5 (0.149) \\
MAC&11463.5 (0.381)&10499 (0.746)&9622.5 (0.166) \\
WEB&10515 (0.57)&10684 (0.934)&9633 (0.149) \\
LOC&9931.5 (0.208)&10580 (0.735)&9464.5 (0.127) \\
ACTIV&10032 (0.365)&10190.5 (0.467)&9649 (0.238) \\
PERSO&10598 (0.666)&11600 (0.256)&9771 (0.224) \\
DIET&11079 (0.697)&11456.5 (0.358)&9902.5 (0.359) \\
\bottomrule
\end{tabular}
\begin{tablenotes}
\small
\item Notes. *$p \le 0.001$
\end{tablenotes}
\end{table}

\begin{table}[!htb]
\centering
\caption{U statistic results of multiple pairwise Mann-Whitney U tests to compare self-reported attitudes toward personal data sharing between European and South Africa participants for different personal data sharing purposes}
\small
\label{tab:pair_eu_sa_personal_data_sharing}
\begin{tabular}{c c c c}
\toprule
PDT & \multicolumn{3}{c}{U Statistic (p-value)}\\
\midrule
& PURPOSE 1 & PURPOSE 2 & PURPOSE 3\\
\midrule
NAME&8686.5 (0.031)&9409 (0.366)&8915.5 (0.139)\\
GEND&9254 (0.184)&8843 (0.086)&8782.5 (0.094)\\
DOB&9565.5 (0.409)&9742 (0.671)&10157 (0.752)\\
NATI&9812.5 (0.636)&9635.5 (0.564)&9140 (0.252)\\
HOME&11603.5 (0.039)&10693 (0.302)&11803 (0.007)\\
EMAIL&9116.5 (0.161)&8878 (0.097)&9223 (0.374)\\
PHONE&8988 (0.119)&9150.5 (0.256)&9802 (0.849)\\
HEIG&8926 (0.088)&8319 (0.013)&10350.5 (0.552)\\
WEIG&8697.5 (0.049)&8485 (0.024)&10517.5 (0.402)\\
SEXO&8851.5 (0.089)&8128 (0.01)&9636 (0.774)\\
ETHN&9161.5 (0.168)&7856 (0.005)&9453 (0.529)\\
RELIG&9143.5 (0.179)&8375 (0.018)&9325.5 (0.416)\\
EDUC&9842.5 (0.788)&9413.5 (0.44)&10472 (0.356)\\
PROF&9997.5 (0.845)&8811 (0.152)&10507 (0.374)\\
CRED&10419 (0.579)&9605.5 (0.499)&11299 (0.031)\\
BANK&10058 (0.91)&9354 (0.311)&11259.5 (0.034)\\
CONT&7428 (*)&7633 (*)&10724 (0.222)\\
PASS&9349 (0.268)&8906.5 (0.115)&11403 (0.018)\\
DRIV&9954.5 (0.851)&9711.5 (0.737)&11523.5 (0.018)\\
VRN&10140 (0.997)&10032.5 (0.992)&11942 (0.002)\\
MEDIC&8451.5 (0.019)&9207 (0.206)&11089 (0.095)\\
FACE&8115 (0.004)&8613 (0.047)&10731 (0.191)\\
VOICE&7892.5 (0.001)&7736.5 (0.001)&10605 (0.331)\\
SOCIA&9529.5 (0.376)&8779.5 (0.068)&11216 (0.053)\\
IP&11439.5 (0.053)&9881 (0.991)&11915.5 (0.002)\\
MAC&10744 (0.329)&9603 (0.588)&12069 (0.001)\\
WEB&10259.5 (0.857)&9786 (0.742)&11478.5 (0.024)\\
LOC&10233 (0.653)&9990.5 (0.938)&11824.5 (0.003)\\
ACTIV&9567.5 (0.75)&9633.5 (0.612)&11237.5 (0.021)\\
PERSO&9020 (0.116)&7982 (0.003)&10177 (0.731)\\
DIET&7271 (*)&8441.5 (0.021)&8433 (0.036)\\
\bottomrule
\end{tabular}
\begin{tablenotes}
\small
\item Notes. *$p \le 0.001$
\end{tablenotes}
\end{table}

\subsubsection{Determining the main effect}

To examine \textbf{H1} and \textbf{H2}, we measured the impact of PDT and PURPOSE on our participants' self-reported privacy attitudes. Considering the significantly different behaviours exhibited by female and male participants, we introduce the gender factor to the investigation to make it more comprehensive. Due to the non-normality of the collected datasets, we adopted the Aligned Rank Transform (ART) methodology to transform the non-parametric data and then applied a three-way ANOVA with Type \( ||| \) sum of squares (SS) (implemented using the Python package \texttt{statsmodels.api.anova\_lm(model, typ=3)}) to conduct the analysis. Type \( ||| \) SS was selected as it is appropriate for unbalanced data and considering all factors by accounting for all interactions, which fits better for our case compared with Type \( || \) SS and Type \( | \) SS.

The dependent variable is C. The gender (G) factor contains 2 distinct values (one for male and another one for female). The factor PDT comprises 31 distinct values, each representing a specific type of personal data. Factor PURPOSE consists of 3 levels, corresponding to PURPOSE 1, PURPOSE 2, and PURPOSE 3 which have been mentioned earlier. Table~\ref{tab:anova} presents the results of the three-way ANOVA for the transformed data, where the $p$-value associated with the intercept term is less than 0.001, suggesting that the model explains a significant portion of the variance. 
In addition, the results indicate that both personal data type (PDT) ($F(30, 27125) = 37.194, p \leq 0.001$) and personal data sharing purpose (PURPOSE) ($F(2, 27125) = 49.113, p \leq 0.001$) have highly significant main effects on the dependent variable (i.e., comfortableness of sharing personal data). This suggests that both factors independently contribute to explaining the variance in the dependent variable, leading to the acceptance of both \textbf{H1} and \textbf{H2}. In addition, a modest and less significant effect of G ($F(1, 27125) = 4.210, p = 0.04$) compared to PDT and PURPOSE is also identified.

\begin{table}[!htb]
\caption{ANOVA results, where df represents `degree of freedom'}
\label{tab:anova}
\small
\begin{tabular}{cccc}
\toprule
\textbf{Factor}  & \textbf{df} & \textbf{F} & \textbf{$p$-value}\\
\midrule
Intercept & 1.0 & 354.185 & *\\
PDT & 30.0 & 37.194 & *\\
PURPOSE & 2.0 & 49.112 & *\\
G & 1.0 & 4.209 & 0.04\\
PDT x PURPOSE & 60.0 & 7.118 & *\\
PDT x G & 30.0 & 1.301 & 0.124\\
PURPOSE x G & 2.0 & 9.162 & *\\
PDT x PURPOSE x G &  60.0 & 3.143 & *\\
Residual & 27125.0 & &\\
\bottomrule
\end{tabular}
\begin{tablenotes}
\small
\item Notes: * = $p \le 0.001$.
\end{tablenotes}
\end{table}

The analysis revealed that the interaction between PDT and PURPOSE is highly significant ($F(60, 27125) = 7.119, p \leq 0.001$). This indicates that the effect of personal data type on the dependent variable varies significantly across different personal data sharing purposes. Moreover, the interaction between PURPOSE and G is also significant ($F(2, 27125) = 9.163, p \leq 0.001$), suggesting that the influence of personal data sharing purpose on the dependent variable is moderated by gender. Furthermore, the three-way interaction among PDT, PURPOSE, and G is significant ($F(60, 27125) = 3.143, p \leq 0.001)$, implying that gender would affect how personal data type and personal data sharing purpose collectively impacts the dependent variable. Although the main effects are identified from the contribution of PDT and PURPOSE, the analysis of interaction effects highlights the complex interplay between these three factors in determining how comfortable participants feel about personal data sharing.

\begin{table}[!htb]
\centering
\begin{threeparttable}
\caption{Results of multiple Wilcoxon Signed Ranks tests}
\label{tab:pair_wsrt}
\small
\begin{tabular}{c c c c}
\toprule
PDT & \multicolumn{3}{c}{Wilcoxon Signed Ranks statistics ($p$-value)}\\
\midrule
& C1 & C2 & C3\\
\midrule
NAME & 753 (*) & 2363 (*) & 6133 (0.360)\\
GEND & 488.5 (*) & 1955 (*) & 5619 (0.718)\\
DOB & 877 (*) & 3671 (*) & 6531.5 (0.367)\\
NATI & 677 (*) & 2627.5 (*) & 6851 (0.897)\\
HOME & 2530 (*) & 8537.5 (0.557) & 4070 (*)\\
EMAIL & 787.5 (*) & 4014.5 (*) & 5531 (0.005)\\
PHONE & 2624 (*) & 7438 (0.017) & 5210.5 (*)\\
HEIG & 1407 (*) & 5340.5 (0.129) & 3863 (*)\\
WEIG & 1117.5 (*) & 4845 (0.055) & 3462 (*)\\
SEXO & 705.5 (*) & 4268.5 (0.098) & 2497 (*)\\
ETHN & 783 (*) & 6613.5 (0.908) & 3271 (*)\\
RELIG & 727.5 (*) & 5176.5 (0.239) & 2697 (*)\\
EDUC & 794 (*) & 5888.5 (0.997) & 3340 (*)\\
PROF & 1490 (*) & 5837.5 (0.813) & 3838.5 (*)\\
CRED & 1609.5 (*) & 6595 (0.156) & 3168.5 (*)\\
BANK & 2092.5 (*) & 5152 (0.006) & 2809 (*)\\
CONT & 853.5 (*) & 4632.5 (0.003) & 2079 (*)\\
PASS & 1401.5 (*) & 8218 (0.173) & 3857.5 (*)\\
DRIV & 2258.5 (*) & 8397 (0.683) & 4823 (*)\\
VRN & 1428 (*) & 5976.5 (0.044) & 3188 (*)\\
MEDIC & 1364.5 (*) & 6130.5 (0.007) & 2778 (*)\\
FACE & 1242 (*) & 6091 (0.021) & 2684.5 (*)\\
VOICE & 903.5 (*) & 4400 (*) & 2002 (*)\\
SOCIA & 1120.5 (*) & 3456 (*) & 2158.5 (*)\\
IP & 600 (*) & 4028.5 (*) & 1708 (*)\\
MAC & 577.5 (*) & 3637 (*) & 2034.5 (*)\\
WEB & 416.5 (*) & 3030.5 (*) & 1452.5 (*)\\
LOC & 873 (*) & 3811.5 (*) & 1882.5 (*)\\
ACTIV & 683 (*) & 3329.5 (*) & 1675.5 (*)\\
PERSO & 1496 (*) & 5498 (*) & 3663 (*)\\
DIET & 762 (*) & 4817.5 (*) & 4353.5 (*)\\
\bottomrule
\end{tabular}
\begin{tablenotes}
\small
\item Notes: * = $p \le 0.001$; \textbf{C1} is for PURPOSE 1 vs.\ PURPOSE 2;
\textbf{C2} is for PURPOSE 1 vs.\ PURPOSE 3; \textbf{C3} is for PURPOSE 2 vs.\ PURPOSE 3.
\end{tablenotes}
\end{threeparttable}
\end{table}

To gain further insights into how travellers' attitudes vary with the identified main effect factors, we performed a series of post-hoc analyses. Given that the analysis involves comparing attitudes towards sharing different types of personal data from the same participants across various stages of travel, we employed Wilcoxon Signed Ranks tests to compare the non-parametric data between these paired groups. Multiple tests were performed to compare the statistical difference between PURPOSE 1 and PURPOSE 2 (i.e., C1 in Table~\ref{tab:pair_wsrt}), PURPOSE 1 and PURPOSE 3 (i.e., C2 in Table~\ref{tab:pair_wsrt}, and PURPOSE 2 and PURPOSE 3 (i.e., C3 in Table~\ref{tab:pair_wsrt}).

As illustrated in Table~\ref{tab:pair_wsrt}, participants demonstrated significantly divergent attitudes towards the majority of PDTs for C1 and C3, while exhibiting consistency in their perspectives towards over half of the PDTs for C2. Specifically, in C1, participants expressed significantly varying attitudes towards sharing all PDTs. Slight differently, within C3, participants displayed similar opinions regarding the sharing of name (NAME), gender (GEND), date of birth (DOB), nationality (NATI), and email address (EMAIL), while exemplifying significant disparities in their attitudes towards other PDTs. Comparatively, within C2, results demonstrated that participants had greater consistency in their comfort level regarding sharing PDTs compared to results for C1 and C3. Notably, participants showcased significantly different attitudes towards approximately half of PDTs, including name (NAME), gender (GEND), date of birth (DOB), nationality (NATI), email address (EMAIL), voice sample (VOICE), social media profile data (SOCIA), Internet Protocol (IP) address (IP), Media Access Control (MAC) address (MAC), web browsing records) (WEB), real time location data (e.g., GPS data) (LOC), activity sensor data (ACTIV), personal interests (PERSO), and dietary requirements (DIET).

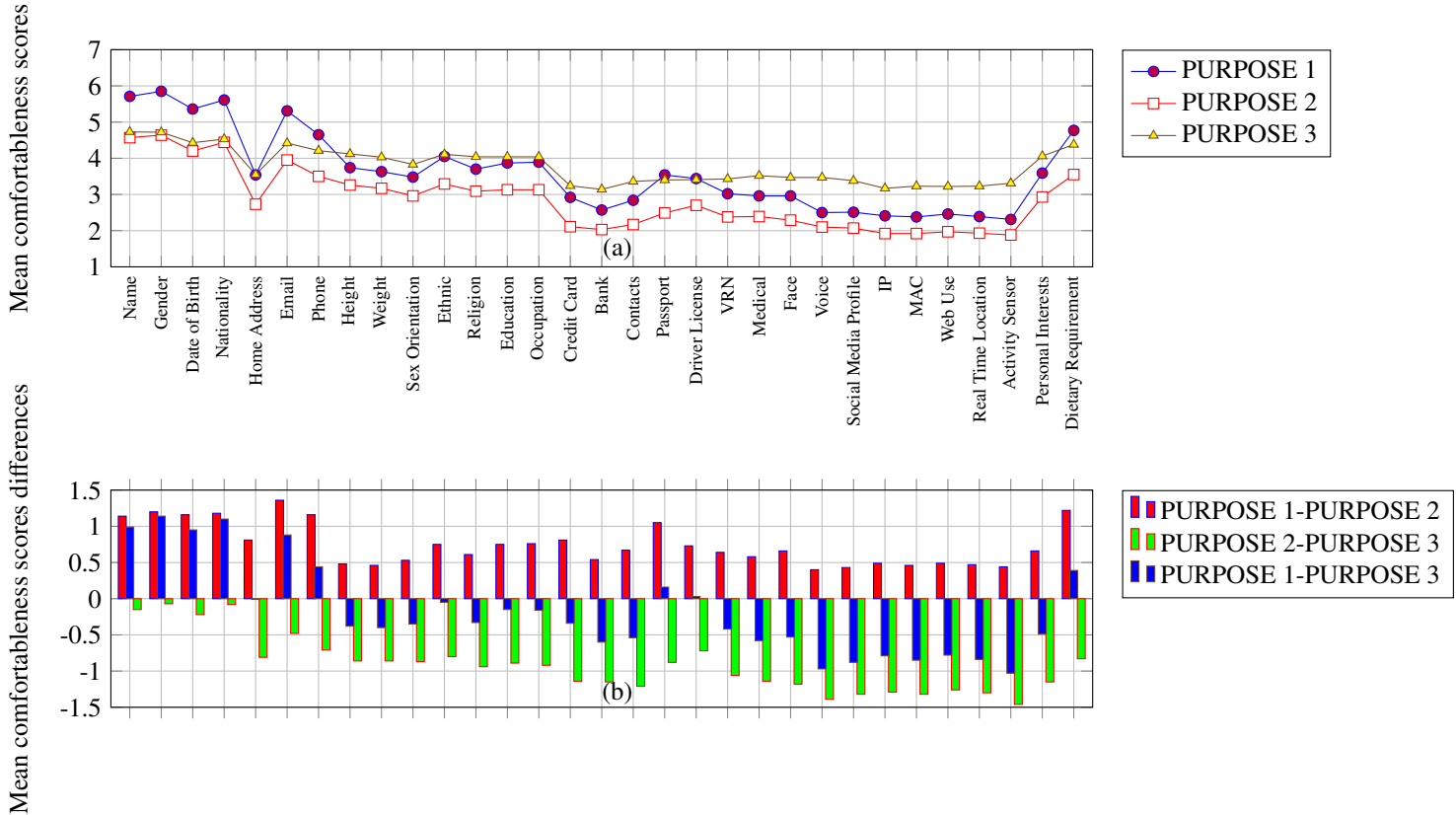
\begin{figure*}[!htb]
\caption{Comparison of average comfortableness scores and comfortableness scores differences}
\label{fig:attitudes_shift}
\centering
\begin{tikzpicture}
\begin{groupplot}[
    group style={
        group size=1 by 2,
        vertical sep=3 cm,
        xticklabels at=edge bottom,
        yticklabels at=edge left,
        horizontal sep=0cm,
    },
    width=0.8\linewidth,
    height=4.5cm,
    grid=major,
    xtick=data,
    x tick label style={rotate=90, anchor=east},
    ymin=-2,
    ymax=8,
    xmin=1,
    xmax=31,
    enlarge x limits=0.02,
    axis on top=false,
]

\nextgroupplot[      
    ymin=1,
    ymax=7,
    ytick={1,2,3,4,5,6,7},
    xticklabels={
        Name, Gender, Date of Birth, Nationality, Home Address, Email, Phone, Height, Weight,
        Sex Orientation, Ethnic, Religion, Education, Occupation, Credit Card, Bank,
        Contacts, Passport, Driver License, VRN, Medical, Face, Voice,
        Social Media Profile, IP, MAC, Web Use, Real Time Location, Activity Sensor,
        Personal Interests, Dietary Requirement
    },
    x tick label style={rotate=0, anchor=east, font=\fontsize{7}{8}\selectfont},
    ylabel={Mean comfortableness scores},
    legend entries={PURPOSE 1, PURPOSE 2, PURPOSE 3},
    legend pos=outer north east,
    legend style={cells={anchor=west}},
    ]
\node at (axis cs:16.5, 1.5) {(a)};

\addplot+[mark=*, mark options={fill=purple}] coordinates {
    (1,5.71) (2,5.85) (3,5.36) (4,5.61) (5,3.54) (6,5.31) (7,4.65) (8,3.74) (9,3.63)
    (10,3.48) (11,4.05) (12,3.70) (13,3.87) (14,3.89) (15,2.92) (16,2.57) (17,2.84)
    (18,3.54) (19,3.44) (20,3.02) (21,2.96) (22,2.96) (23,2.50) (24,2.51) (25,2.41)
    (26,2.38) (27,2.46) (28,2.39) (29,2.31) (30,3.59) (31,4.77)
};
\addplot+[mark=square*, mark options={fill=white}] coordinates {
    (1,4.57) (2,4.64) (3,4.20) (4,4.44) (5,2.73) (6,3.95) (7,3.50) (8,3.26) (9,3.17)
    (10,2.96) (11,3.29) (12,3.09) (13,3.13) (14,3.13) (15,2.11) (16,2.03) (17,2.17)
    (18,2.49) (19,2.70) (20,2.38) (21,2.39) (22,2.29) (23,2.10) (24,2.07) (25,1.92)
    (26,1.92) (27,1.97) (28,1.93) (29,1.88) (30,2.93) (31,3.55)
};
\addplot+[mark=triangle*, mark options={fill=yellow}] coordinates {
    (1,4.73) (2,4.72) (3,4.43) (4,4.53) (5,3.55) (6,4.42) (7,4.21) (8,4.12) (9,4.03)
    (10,3.83) (11,4.11) (12,4.04) (13,4.04) (14,4.04) (15,3.24) (16,3.14) (17,3.36)
    (18,3.40) (19,3.41) (20,3.43) (21,3.52) (22,3.47) (23,3.47) (24,3.38) (25,3.17)
    (26,3.23) (27,3.22) (28,3.23) (29,3.31) (30,4.06) (31,4.38)
};

\nextgroupplot[
    ymin=-1.5,
    ymax=1.5,
    ytick={-1.5,-1,-0.5,0,0.5,1,1.5},
    yticklabels={-1.5,-1,-0.5,0,0.5,1,1.5},
    xticklabels={},
    ylabel={Mean comfortableness scores differences},
    legend pos=outer north east,
    legend style={cells={anchor=west}},
    ybar,
    bar width=3pt,
]
\node at (axis cs:16.5, -1.3) {(b)};
\addplot+[ybar, bar shift=-0.1cm, fill=red, no markers] coordinates {
    (1,1.14) (2,1.20) (3,1.16) (4,1.18) (5,0.81) (6,1.36) (7,1.16) (8,0.48) (9,0.46)
    (10,0.53) (11,0.75) (12,0.61) (13,0.75) (14,0.76) (15,0.81) (16,0.54) (17,0.67)
    (18,1.05) (19,0.73) (20,0.64) (21,0.58) (22,0.66) (23,0.40) (24,0.43) (25,0.49)
    (26,0.46) (27,0.49) (28,0.47) (29,0.44) (30,0.66) (31,1.22)
};
\addplot+[ybar, bar shift=0.1cm, fill=green, no markers] coordinates {
    (1,-0.15) (2,-0.07) (3,-0.22) (4,-0.08) (5,-0.81) (6,-0.48) (7,-0.71) (8,-0.86) (9,-0.86)
    (10,-0.87) (11,-0.80) (12,-0.94) (13,-0.89) (14,-0.92) (15,-1.14) (16,-1.15) (17,-1.21)
    (18,-0.88) (19,-0.72) (20,-1.06) (21,-1.14) (22,-1.18) (23,-1.39) (24,-1.32) (25,-1.29)
    (26,-1.32) (27,-1.26) (28,-1.30) (29,-1.46) (30,-1.15) (31,-0.83)
};
\addplot+[ybar, bar shift=0cm, fill=blue, no markers] coordinates {
    (1,0.99) (2,1.14) (3,0.95) (4,1.10) (5,-0.01) (6,0.88) (7,0.44) (8,-0.38) (9,-0.40)
    (10,-0.35) (11,-0.05) (12,-0.33) (13,-0.15) (14,-0.16) (15,-0.34) (16,-0.60) (17,-0.54)
    (18,0.16) (19,0.03) (20,-0.42) (21,-0.58) (22,-0.53) (23,-0.97) (24,-0.88) (25,-0.79)
    (26,-0.85) (27,-0.78) (28,-0.84) (29,-1.03) (30,-0.49) (31,0.39)
};
\addlegendimage{area legend,fill=blue,draw=blue}
\addlegendentry{PURPOSE 1-PURPOSE 2}
\addlegendimage{area legend,fill=green,draw=green}
\addlegendentry{PURPOSE 2-PURPOSE 3}
\addlegendimage{area legend,fill=red,draw=red}
\addlegendentry{PURPOSE 1-PURPOSE 3}
\end{groupplot}
\end{tikzpicture}
\end{figure*}

To further facilitate the interpretation of the results, we 1) plot the mean comfortableness scores for different settings as illustrated in Figure~\ref{fig:attitudes_shift} (a); and 2) perform subtractions of comfortableness scores for all personal data types across different settings, and produce a bar chart as depicted in Figure~\ref{fig:attitudes_shift} (b), aiming to observe how participants shift their attitudes when the intended purpose of personal data sharing changes. In Figure~\ref{fig:attitudes_shift} (b), a positive value indicates that participants are less comfortable sharing their personal data for one intended purpose than another, while a negative value suggests the opposite. For the sake of convenience and consistency, we will refer to the former as a negative attitude shift and the latter as a positive attitude shift.

Across all PURPOSEs, participants exhibited a higher level of comfort when sharing details such as name, gender, date of birth, nationality, email address, and phone number, in contrast to other types of personal data, as depicted in Figure~\ref{fig:attitudes_shift} (a). This may initially seem counter-intuitive, as these data are typically viewed as identifiable and sensitive. However, travellers are accustomed to providing such information to complete holiday bookings, especially for international travel. This is also somewhat supported by findings in another study about privacy in smart tourism~\cite{Francisco-F2022}, which indicates that information that can be effortlessly obtained is often not perceived as `too personal'. Moreover, finance data (e.g., credit card, bank information), biometric data (e.g., facial images and voice samples), and behavioural data (e.g., web usage data, real-time location data, and activity sensor data) are regarded as more sensitive data that participants were less comfortable sharing across different PURPOSEs. 

As depicted in Figure~\ref{fig:attitudes_shift} (a), PURPOSE 2 has lower values across all personal data types compared to PURPOSE 1, resulting in negative attitude shifts between two sharing purposes, visualised by the red bars in Figure~\ref{fig:attitudes_shift} (b). This suggests that participants were less comfortable (i.e., more sensitive) about sharing personal data to entities (i.e., third-party service providers, subsidiaries, parent companies) that are associated with the booking sites participants used for travel booking (i.e., PURPOSE 1). Similarly, when comparing the change of sharing behaviour from PURPOSE 2 to PURPOSE 3 as illustrated as green bars in Figure~\ref{fig:attitudes_shift} (b), results suggest that participants exhibited positive attitude shifts across all personal data types, meaning that participants were more comfortable sharing personal data during travels for booking various activities and services compared with PURPOSE 2. As reported in another study~\cite{Yuan-H2023}, the scope and scale of sharing personal data with such affiliated entities are often ambiguously outlined or obscured within the corresponding privacy policies, suggesting a lack of transparency regarding how and to what extent the personal data would be shared beyond the initial travel booking sites. In addition, as the data sharing for PURPOSE 2 is governed by the booking sites themselves, travellers have little control. Consequently, the sense of detachment or loss of control could indeed be a contributing factor to the low levels of comfort experienced by travellers compared to other PURPOSEs.

The comparison between PURPOSE 1 and PURPOSE 3 as demonstrated as blue bars in Figure~\ref{fig:attitudes_shift} is essentially the comparison of sharing behaviour between the `pre-travel' stage and `during travel' stage. For name, gender, date of birth, nationality, email address, phone number, and dietary requirements, participants exhibited negative attitude shifts, meaning that participants were less comfortable sharing these data. We would argue that this might be related to the trust in the booking sites at different stages. As reported in the study~\cite{Ioannou-A2020}, trust is negatively associated with traveller's online privacy concerns. People would spend more time planning and booking their holidays at the pre-travel stage, leading to a higher level of familiarity and trust with the booking sites they frequently visit. Whereas, activity bookings and service bookings during holidays/travels are often more spontaneous and more benefits-driven/orientated.

Moreover, based on findings from the same study~\citet{Ioannou-A2020}, travellers' perceived expected benefits can outweigh privacy concerns in their privacy decisions regarding their biometric data (e.g., facial images and voice samples) and behavioural data (e.g., personal preferences, web usage data, real-time location data, and activity sensor data). This could partially explain why we could see relatively large positive shifts of attitudes towards sharing personal data types such as face image, voice sample, Web-usage data (i.e., web browsing records), real-time location data (e.g., GPS data), and activity sensor data in Figure~\ref{fig:attitudes_shift} (b).

Overall, the differences observed between Figure~\ref{fig:attitudes_shift} (a) and \ref{fig:attitudes_shift} (b) across PURPOSE 1, PURPOSE 2 and PURPOSE 3 echo the results of the statistical analysis presented above, suggesting that travellers have different privacy perceptions/attitudes towards personal data sharing for the different intended purpose at different stages of leisure travel. This could be explained by the finding identified by~\citet{Masur-P2019} and \citet{Acquisti-A2015}, who argued that the privacy levels, privacy perceptions, and cost-benefit trade-offs in privacy decisions are context-dependent and change across situations. Individuals might change their behaviour from very concerned to complete apathy. To this end, we recommend that more studies that consider different contexts need to be conducted to further confirm and explain the observations identified in this survey study.

\subsection{Leisure travellers' perception of the travel experience sharing using social media platforms at different stages of leisure travel}
\label{sec:sharing}

In this part, our focus is to investigate the use of social media platforms for sharing content in different stages of leisure travel. To establish background information about travellers' knowledge and usage towards some of the popular social media platforms in general, we studied the general usage of selected social media platforms, which has been reported in Section~\ref{sec:general_use} (see Table~\ref{tab:general_usage_social}). Notably, Facebook, Instagram, and YouTube emerge as the favoured platforms among participants, while Pinterest and Snapchat are comparatively less popular. Furthermore, to examine \textbf{H3} and \textbf{H4}, we asked participants more specifically about their sharing behaviour using social media platforms pre-travel, during travel, and after travel respectively, where the options provided for participants to choose from include `Share with friends to keep updated', `Share with family members to keep updated', `Share with colleagues', `Share publicly', and `Share privately to myself to keep a record', `Do not share', and `Prefer not to say'. These options are derived based on the observations of how these social media platforms support content sharing. For instance, the default audience for sharing on Facebook includes `Public', `Friends', `Specific friends', `Only me', `Custom', etc. Instagram can achieve similar functionalities by manipulating the settings such as switching on/off `private account' and selecting `Close friends'.

\begin{table}[!htb]
\centering
\caption{Results of multiple Chi-square tests to compare self-reported sharing behaviour using social media platforms between different travel stages}
\label{tab:cst_sharing_travel_stages}
\small
\begin{tabular}{c c c c}
\toprule
Social Media & \multicolumn{3}{c}{Chi-square test's statistic ($p$-value)}\\
\midrule
& Pre- vs. During & Pre- vs.\ After  &  During vs.\ After\\
\midrule
Facebook & 15.805 (0.007) & 33.834 (*) & 5.788 (0.327)\\
Instagram & 10.467 (0.063) & 22.775 (*) & 3.677 (0.597)\\
Twitter & 6.636 (0.249) & 11.158 (0.048) & 3.900 (0.564)\\
TikTok & 3.024 (0.696) & 4.429 (0.489) & 1.685 (0.891)\\
Snapchat & 5.733 (0.333) & 8.092 (0.151) & 0.877 (0.972)\\
YouTube & 11.485 (0.043) & 10.573 (0.061) & 1.278 (0.937)\\
Pinterest & 4.820 (0.438) & 2.563 (0.767) & 1.283 (0.937)\\
\bottomrule
\end{tabular}
\begin{tablenotes}
\item \small Notes: * = $p \le 0.001$.
\end{tablenotes}
\end{table}

\begin{figure}[!htb]
\centering
\caption{Participants' self-reported sharing behaviour for three travel stages: a) pre-travel; b) during- travel; and c) after-travel}
\label{fig:holiday_share}
\begin{subfigure}{\linewidth}
\centering
\begin{tikzpicture}
\begin{axis}[
    xbar stacked, 
    bar width=7.5pt, 
    legend style={at={(0.5,-0.1)}, anchor=north, legend columns=2}, 
    xlabel={(a) Sharing percentage}, 
    symbolic y coords={Facebook, Instagram, Twitter, Pinterest, Snapchat, YouTube, TikTok}, 
    ytick=data, 
    xmin=0, 
    xmax=1, 
    enlarge y limits=0.1, 
    xticklabel={\pgfmathparse{\tick*100}\pgfmathprintnumber{\pgfmathresult}\%}, 
    width=7cm, 
    height=3.9cm, 
    ]
\addplot+[xbar, draw=blue!50, fill=blue!50] plot coordinates {(0.259,Facebook) (0.299,Instagram) (0.110,Twitter) (0.028,Pinterest) (0.075,Snapchat) (0.067,YouTube) (0.097,TikTok)};
\addplot+[xbar, draw=red!50, fill=red!50] plot coordinates {(0.227,Facebook) (0.181,Instagram) (0.063,Twitter) (0.006,Pinterest) (0.048,Snapchat)  (0.038,YouTube) (0.052,TikTok)};
\addplot+[xbar, draw=yellow, fill=yellow] plot coordinates {(0.066,Facebook) (0.067,Instagram) (0.027,Twitter) (0.006,Pinterest) (0.021,Snapchat) (0.026,YouTube) (0.026,TikTok)};
\addplot+[xbar, draw=gray, fill=gray] plot coordinates {(0.079,Facebook) (0.118,Instagram) (0.096,Twitter) (0.034,Pinterest) (0.018,Snapchat) (0.055,YouTube) (0.066,TikTok)};
\addplot+[xbar, draw=purple!50, fill=purple!50] plot coordinates {(0.076,Facebook) (0.083,Instagram) (0.052,Twitter) (0.040,Pinterest) (0.042,Snapchat) (0.055,YouTube) (0.037,TikTok)};
\addplot+[xbar, draw=green, fill=green]  plot coordinates {(0.293,Facebook) (0.252,Instagram) (0.652,Twitter) (0.885,Pinterest) (0.797,Snapchat) (0.758,YouTube) (0.722,TikTok)};
\end{axis}
\end{tikzpicture}
\end{subfigure}

\begin{subfigure}{\linewidth}
\centering
\begin{tikzpicture}
\begin{axis}[
    xbar stacked, 
    bar width=7.5pt, 
    legend style={at={(0.5,-0.1)}, anchor=north, legend columns=2}, 
    xlabel={(b) Sharing percentage}, 
    symbolic y coords={Facebook, Instagram, Twitter, Pinterest, Snapchat, YouTube, TikTok}, 
    ytick=data, 
    xmin=0, 
    xmax=1, 
    enlarge y limits=0.1, 
    xticklabel={\pgfmathparse{\tick*100}\pgfmathprintnumber{\pgfmathresult}\%}, 
    width=7cm, 
    height=3.9cm, 
    ]
\addplot+[xbar, draw=blue!50, fill=blue!50] plot coordinates {(0.220,Facebook) (0.263,Instagram) (0.094,Twitter) (0.025,Pinterest) (0.075,Snapchat) (0.046,YouTube) (0.081,TikTok)};
\addplot+[xbar, draw=red!50, fill=red!50] plot coordinates {(0.178,Facebook) (0.153,Instagram) (0.046,Twitter) (0.019,Pinterest) (0.022,Snapchat) (0.025,YouTube) (0.030,TikTok)};
\addplot+[xbar, draw=yellow, fill=yellow] plot coordinates {(0.056,Facebook) (0.064,Instagram) (0.037,Twitter) (0.006,Pinterest) (0.013,Snapchat) (0.006,YouTube) (0.027,TikTok)};
\addplot+[xbar, draw=gray, fill=gray] plot coordinates {(0.063,Facebook) (0.096,Instagram) (0.095,Twitter) (0.028,Pinterest) (0.009,Snapchat) (0.046,YouTube) (0.066,TikTok)};
\addplot+[xbar, draw=purple!50, fill=purple!50] plot coordinates {(0.063,Facebook) (0.076,Instagram) (0.023,Twitter) (0.019,Pinterest) (0.031,Snapchat) (0.028,YouTube) (0.033,TikTok)};
\addplot+[xbar, draw=green, fill=green] plot coordinates {(0.419,Facebook) (0.348,Instagram) (0.705,Twitter) (0.902,Pinterest) (0.850,Snapchat) (0.850,YouTube) (0.764,TikTok)};
\end{axis}
\end{tikzpicture}
\end{subfigure}

\begin{subfigure}{\linewidth}
\centering
\begin{tikzpicture}
\begin{axis}[
    xbar stacked, 
    bar width=7.5pt, 
    legend style={at={(0.5,-0.5)}, anchor=north, legend columns=2, legend cell align=left}, 
    xlabel={(c) Sharing percentage}, 
    symbolic y coords={Facebook, Instagram, Twitter, Pinterest, Snapchat, YouTube, TikTok}, 
    ytick=data, 
    xmin=0, 
    xmax=1, 
    enlarge y limits=0.1, 
    xticklabel={\pgfmathparse{\tick*100}\pgfmathprintnumber{\pgfmathresult}\%}, 
    width=7cm, 
    height=3.9cm, 
    ]
\addplot+[xbar, draw=blue!50, fill=blue!50] plot coordinates {(0.220,Facebook) (0.263,Instagram) (0.095,Twitter) (0.025,Pinterest) (0.075,Snapchat) (0.046,YouTube) (0.081,TikTok)};
\addplot+[xbar, draw=red!50, fill=red!50] plot coordinates {(0.178,Facebook) (0.153,Instagram) (0.046,Twitter) (0.019,Pinterest) (0.022,Snapchat) (0.025,YouTube) (0.030,TikTok)};
\addplot+[xbar, draw=yellow, fill=yellow] plot coordinates {(0.056,Facebook) (0.064,Instagram) (0.037,Twitter) (0.006,Pinterest) (0.013,Snapchat) (0.006,YouTube) (0.027,TikTok)};
\addplot+[xbar, draw=gray, fill=gray] plot coordinates {(0.063,Facebook) (0.096,Instagram) (0.095,Twitter) (0.028,Pinterest) (0.009,Snapchat) (0.046,YouTube) (0.066,TikTok)};
\addplot+[xbar, draw=purple!50, fill=purple!50] plot coordinates {(0.063,Facebook) (0.076,Instagram) (0.023,Twitter) (0.019,Pinterest) (0.031,Snapchat) (0.028,YouTube) (0.033,TikTok)};
\addplot+[xbar, draw=green, fill=green] plot coordinates {(0.419,Facebook) (0.348,Instagram) (0.705,Twitter) (0.902,Pinterest) (0.850,Snapchat) (0.850,YouTube) (0.764,TikTok)};
\legend{Share with friends, Share with family, Share with colleagues, Share with the public, Share with myself, Do not share}
\end{axis}
\end{tikzpicture}
\end{subfigure}
\end{figure}
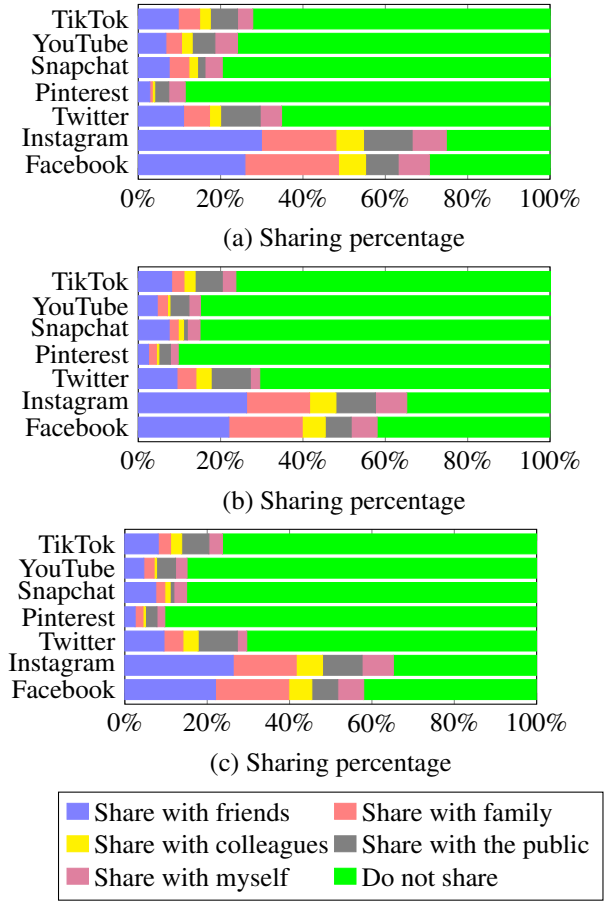

We started the analysis by aggregating participants' votes for each sharing option across various social media platforms. We used Chi-square tests to compare the aggregated data between different travel stages. As shown in Table~\ref{tab:cst_sharing_travel_stages}, participants generally displayed consistent behaviours across the selected social media platforms throughout various travel stages. The notable exceptions are the differences observed in how individuals reported about sharing content on Facebook and Instagram between the pre-travel and after-travel stages. To further facilitate the analysis, we produced stacked bar charts to visually inspect participants' self-reported sharing behaviours for the three travel stages. As illustrated in Figures~\ref{fig:holiday_share} (a) and (c), nearly 80\% of participants would use Instagram to share their travel-related content to different audiences, and around 70\% of participants exhibited a similar sharing behaviour of using Facebook. In comparison, the percentages fall below 70\% and 60\% for Instagram and Facebook, respectively, at the after-travel stage. Nevertheless, we can still observe relatively bigger areas corresponding to `Share Family' and `Share Friends' for Facebook and Instagram across three stacked bar charts. This indicates that participants prefer creating and sharing leisure travel-related content with their friends and family members across different stages of leisure travel using Facebook and Instagram while showing less enthusiasm for sharing via other selected social media platforms. Apart from Instagram and Facebook, we can see similar patterns across different stages. A considerable portion of participants indicated that they normally do not share content on TikTok, YouTube, Snapchat, Pinterest, and Twitter, despite the widespread popularity of these social media platforms. This trend is particularly notable for Twitter, as it was reported to be used by participants multiple times per day, as shown in Table~\ref{tab:general_usage_social} from Section~\ref{sec:general_use}.

According to the well-known theory of the 90-9-1 principle~\cite{Nielsen-J2021}, 90\% of social network users, known as Lurkers, predominantly observe without active participation. 9\% of users are known as Contributors who would contribute content occasionally, while 1\% of the so-called Superusers, generate the vast majority of the content. This has been confirmed empirically in a separate study~\cite{Van-T2014}, which reported that Lurkers, Contributors, and Superusers were responsible for a weighted average of 1.3\%, 24.0\% and 74.7\% of content respectively. Another study~\cite{Brandtzaeg-P2011} indicated that instead of the 90-9-1 principle, a 50-30-20 rule is more representative of small and locally bonded online communities. Given these insights, the observed findings in this study align with the expectations and to some extent highlight participants' tendency towards passive content consumption over active creation and sharing throughout leisure travels.

Furthermore, we investigated the changes in individual participants' sharing preferences across different stages of leisure travel. We calculated the percentage of changes by tallying the number of participants who change their votes for the same sharing intention 1) before and during the leisure travel, and 2) during and after the leisure travel. On average, we found that only 2.8\% ($\sigma=0.028$) of participants changed their sharing preferences in the former case, and only 1.9\% ($\sigma=0.018$) of participants had attitude changes for the latter case.

Based on the findings presented earlier in this paper, it is sensible to introduce previous travel frequency, gender, and country of residence as predictors for predicting if participants would share under different circumstances. We treated the selected social media platforms as one predictor (S-PLAT) with seven categorical values to represent different social media platforms, the leisure travel stage (L-STAGE) as one predictor with three categorical values to represent three different stages, gender (G) as one predictor with two categorical values, previous travel frequency (P-FREQ) as one predictor with two categorical values, country of residence (C-RESI) as one predictor with two categorical values, and the sharing option (S-OPTION) as another predictor that was categorised into seven values representing different options for sharing. It is worth noting that this is also referred to as `sharing intention' for the rest of the analysis. The dependent variable was the traveller's decision to share (D-SHARE), which we encoded as either 1 (Yes) or 0 (No). A multivariate logistic regression model was employed to assess the predictability of the dependent variable (D-SHARE) using six predictors S-PLAT, L-STAGE, G, P-FREQ, C-RESI, and S-OPTION. As reported in Table~\ref{tab:regression}, the calculated log-likelihood ratio (LLR) $p$-value $\le 0.001$, suggesting that the model as a whole is statistically significant. In addition, we also measured the VIF (Variance Inflation Factor) to see how much multicollinearity among independent variables can inflate the variance of the estimated regression coefficients, and the results suggest that multicollinearity would not cause issues in our model.

\begin{table}[!htb]
\centering
\caption{Results of the multivariate logistic regression model}
\label{tab:regression}
\small
\begin{tabular}{lc|lc}
\toprule
\multicolumn{4}{c}{$LL = -9473.6$, $LL\_null = -9959.7$} \\
\multicolumn{4}{c}{$p\_LLR \le .001$, pseudo-$R^2 = 0.05$}\\
\midrule
\textbf{Predictors} & $\beta$ (p-value) & \textbf{Predictors} & $\beta$ (p-value)\\
\midrule
L-STAGE & 0.017 (0.616) & P-FREQ & 0.262 (*)\\
S-PLAT & -0.135 (*) & C-RESI & -0.103 (*)\\
G & 0.114 (*) & S-OPTION & 0.079 (*)\\
intercept & -0.818 (*) & &\\
\bottomrule
\end{tabular}
\begin{tablenotes}
\item \small Notes: * = $p \le 0.001$; LL stands for Log-Likelihood; LLR stands for Log-Likelihood Ratio.
\end{tablenotes}
\end{table}

The coefficient with the $p$-value of L-STAGE indicates that the stage of leisure travel does not significantly impact how people share content on social media platforms. On the contrary, significant effects have been identified for all other predictors. Here, coefficients indicate the direction and strength of the relationship between predictors and the outcome, with positive values suggesting increased odds and negative values decreasing odds, while larger absolute values denote stronger associations. By looking at the coefficients of the predictors that have a significant impact, gender and previous travel frequency positively can predict the dependent variable, whereas social media platforms and country of residence exhibit inverse relationships.

In summary, the findings presented in this section provide support for Hypothesis \textbf{H3}, while evidence for Hypothesis \textbf{H4} remains inconclusive. Nevertheless, consistent with the results outlined in Section~\ref{sec:privacy_attitudes}, previous travel frequency, gender, and country of residence emerged to influence participants' self-reported responses at various levels.

\section{Discussions}
\label{sec:discussion}

In the context of leisure travel, privacy concerns about personal data sharing can be related to two components of leisure travel. One is about the use of personal data to facilitate travel-related bookings, and the other one is about how travellers share their travel-related experience/content on social media platforms. In this study, we aimed to examine potential disparities in leisure travellers' self-reported attitudes and behaviour across different stages of travel from these two perspectives. We discovered that travellers' comfort levels with sharing various types of personal data varied depending on the specific circumstances under which the personal data would be used, suggesting that travellers do have different levels of privacy perceptions towards different personal data sharing in the context of leisure travel bookings. Surprisingly, participants were more relaxed about sharing personal data, including `name', `gender', `date of birth', `nationality', `email address' and `phone number', which are often regarded as sensitive data. This openness likely stems from the acceptance of such data as essential/standard requirements for travel bookings, leading to a normalisation of their collection by booking sites. In addition, information that is readily accessible is often not regarded as overly private~\cite{Francisco-F2022}.

As discussed in Section~\ref{sec:relatedwork}, extensive research has been conducted to investigate how individuals would share travel experiences using social media platforms during leisure travels or holidays. It was assumed that travellers may display distinct sharing behaviours at different stages of their travel journey. However, the findings of this study failed to confirm this assumption, indicating that the travel stage does not significantly influence how people share their travel-related experiences on social media platforms, suggesting a uniform approach to general online privacy and information sharing among travellers. In contrast, other factors such as gender, previous travel frequency, and country of residence have a significant impact on the sharing decision-making process, implying that a less uniform approach should be adopted for people with different demographics. This aligns with the findings from past research conducted by~\citet{Masur-P2019} and \citet{Acquisti-A2015}, highlighting the dynamic nature of privacy perceptions in the context of leisure travel. 

\subsection{Recommendations/contributions to privacy and security community}

This subsection explains some potential applications and contributions to the relevant research community by leveraging and expanding upon the research findings presented in this paper.

\paragraph{\textbf{Privacy nudging for travellers and service providers}} Implementing data protection technologies that provide contextual privacy nudges can enhance user awareness and informed decision-making regarding data sharing during travel. For users, in various travel contexts, appropriate privacy nudges can remind users of potential risks associated with sharing specific types of data at different stages of their journey. For service providers in the travel industry, the insights revealed from this study can facilitate the design of specialised training programmes that emphasise the importance of data protection and offer guidelines on effectively communicating privacy policies to users.

\paragraph{\textbf{Dynamic consent mechanisms for data sharing}} Developing dynamic and adaptive consent mechanisms that can provide context-specific consent options, ensuring that users are fully informed about the type of data they are sharing and the associated risks at different travel stages. For example, users may be presented with more detailed consent requests when sharing sensitive health information during international travel compared to sharing less sensitive data, such as travel photos. Moreover, the findings of the gender-based differences highlight the importance of considering gender when designing such privacy-enhancing technologies, as different user groups may require tailored solutions to address their specific concerns and behaviours. Additionally, implementing stage-specific and real-time consent management systems allows users to easily update their data-sharing preferences as their travel context evolves, ensuring continuous and contextually appropriate consent throughout their journey.

\paragraph {\textbf{Enhanced anonymisation techniques}} Developing context-aware anonymisation techniques can further protect user privacy by tailoring the level of data anonymisation based on how sensitive people would treat specific types of data at different travel stages. For example, a more rigorous anonymization strategy might be employed when sharing data with third parties to offer additional benefits (PURPOSE 2) compared to sharing personal data for holiday bookings (PURPOSE 1), as our findings suggest that travellers tend to have more relaxed attitudes towards the latter. Additionally, such dynamic and adaptive anonymisation can also enable travel service providers to conduct meaningful data analysis to optimise their offerings without compromising user privacy.

\paragraph{\textbf{Adaptive personal data vaults}} Considering the study's findings on the complex and evolving nature of participants' privacy attitudes towards sharing different personal data across different travel stages, it underscores the importance of enhancing the research and development on creating personal data repositories where users can store their travel-related data and control access permissions adaptively. This allows users to manage who can access specific data types and for what purposes, offering a granular level of control over their personal data. Additionally, adaptive identity management can also be introduced to enhance privacy by enabling users to utilise such personal data vaults to authenticate their identities and share verified credentials without revealing unnecessary personal information.

\subsection{Limitations and future work}

\paragraph{\textbf{Limitation of using the survey study}} It is important to acknowledge the primary limitation of this study, which stems from the use of an online survey. Such a methodology can only capture self-reported/declared responses from participants, potentially not fully representing real-world attitudes and behaviours. In addition, a survey study typically can not fully explain the underlying reasons for the observations and insights derived from participants' responses. We would argue that more focus group or interview studies are needed to further consolidate and elaborate the findings identified in this study. Nevertheless, we are developing a mobile app designed to function as a personal data manager, that can inform users about their personal data shared with and collected by selected online services to provide them with insights into their data-sharing activities and what they could do to better protect their own privacy while balancing the need of travel experience. We plan to conduct a field study in the near future, recruiting participants to use this mobile app during a real holiday period. This field study is anticipated to provide data that can further validate and confirm the findings identified in this paper.

\paragraph{\textbf{Limitation of imbalanced sample}} We acknowledge the imbalance in our sample, with over two-thirds of the participants coming from Europe. Our intention was to recruit participants globally, as our inclusion criteria were designed to consider the entire world population. The survey was made available via the Prolific platform to all countries without any intentional preference or intervention to increase participation from specific regions. Despite this, we believe that the data from European and South Africa travellers is still significant, as it provides a relevant and meaningful case study in understanding how individuals from different regions report their attitudes toward privacy and content sharing. We agree that a more balanced global representation would strengthen the generalisability of our findings. We aim to conduct further user studies with a stronger focus on obtaining more diverse and representative samples from various global regions. This will help validate our findings across a broader range of cultural and geographical contexts.

\paragraph{\textbf{Limitation of the exclusion of Instant Messengers}} IMs differ fundamentally from the social media platforms surveyed, as they are typically less suited for public sharing and present distinct privacy concerns. Unlike social media platforms, IMs prioritise private, often end-to-end encrypted, communication between individuals or small groups, which presents unique challenges in terms of privacy and security. Recognising the significance of IMs, we propose that their inclusion deserves a separate study to fully explore these unique aspects and their implications on user privacy and data sharing in the context of leisure travel. As reported in this study, most of our survey participants were from Europe and South Africa, which may limit the generalisability of the findings to the broader global population. While this might be seen as a limitation, the study still demonstrates the influence of regional differences on participants' responses, highlighting the value of the findings. For future research, we recommend adopting a more targeted recruitment strategy focusing on a single region or significantly expanding the participant pool to ensure a more comprehensive representation of diverse populations.

\paragraph{\textbf{Limitation of the existing passive data control and sharing dynamics}} Considering that travellers have limited control over the collection and sharing of their personal data, making them more sensitive to data sharing for different purposes. Even with consent, data sharing is often ``passive'', controlled by firms without transparency for travellers, as noted by~\citet{Wieringa-J2021}. In contrast, travellers are more comfortable with proactively sharing experiences on social media, where perceived vulnerability is lower, and they continue sharing throughout different travel stages. Besides passive and proactive sharing, \citet{Wieringa-J2021} introduced a third approach—agents acting on behalf of individuals. The conceptual model and methods are yet to be tested in practice, in particular in a complex context such as travel. This is identified as another area for future research.

\section{Conclusion}
\label{sec:conclusion}

This paper presents findings from an online survey that aimed at examining privacy attitudes and behaviours of leisure travellers towards the sharing of personal data at different stages of leisure travel. The results showed that both the personal data type and the intended personal data sharing purpose have a significant influence on participants' self-reported comfortableness in sharing their personal data. Additionally, we also identified that participants exhibited significantly different data sharing behaviours on different social media platforms while maintaining more consistent sharing patterns on individual social media platforms across different travel phases. Nevertheless, factors such as gender, previous travel frequency and country of residence of participants play important roles in shaping participants' self-reported privacy attitudes and perceptions to some extent. Although our work has a specific scope (people's data sharing attitudes and behaviours in different stages of leisure travel), the results can help inform the privacy-enhancing technologies community regarding future research on a wider range of topics.

\section*{Acknowledgements}

The authors were supported by the Engineering and Physical Sciences Research Council, UK Research and Innovation (UKRI), as part the project ``PriVELT: PRIvacy-aware personal data management and Value Enhancement for Leisure Travellers'', under the grant numbers EP/R033749/1 and EP/R033609/1.

\bibliographystyle{ACM-Reference-Format}
\bibliography{main}

\section{Appendix}
\label{sec:appendix}

\begin{table*}[!htb]
\caption{A heat-map showing the male participants' self-reported travelling patterns with different travelling companionship types}
\label{tab:male_travel_experience}
\centering
\pgfplotstabletypeset[%
     color cells={min=0,max=100,textcolor=black},
     /pgfplots/colormap={}{rgb255=(255,255,255) rgb255=(255,0,0)},
    /pgf/number format/fixed,
    /pgf/number format/precision=3,
    col sep=comma,
    columns/-/.style={reset styles,string type}%
]{
-, Alone, Colleagues, Friends, Family without kids, Family with kids
Once per week, 8, 3, 5, 1, 6
Once per month, 24, 8, 14, 12, 10
Once per quarter, 15, 16, 40, 25, 24
Once per half year, 10, 12, 36, 31, 37
Once per year, 21, 26, 38, 32, 30
Less than once per year, 70, 64, 27, 45, 30
}
\end{table*}

\begin{table*}[!htb]
\caption{A heat-map showing the female participants' self-reported travelling patterns with different travelling companionship types}
\label{tab:female_travel_experience}
\centering
\pgfplotstabletypeset[%
     color cells={min=0,max=80,textcolor=black},
     /pgfplots/colormap={}{rgb255=(255,255,255) rgb255=(255,0,0)},
    /pgf/number format/fixed,
    /pgf/number format/precision=3,
    col sep=comma,
    columns/-/.style={reset styles,string type}%
]{
-,Alone, Colleagues, Friends, Family without kids, Family with kids
Once per week,12, 3, 5, 6, 8
Once per month,23, 12, 23, 14, 10
Once per quarter,18, 17, 24, 17, 23
Once per half year,15, 16, 26, 18, 17
Once per year,23, 20, 28, 29, 35
Less than once per year,44,	58,	31,	48,	32
}
\end{table*}

\begin{table*}[!htb]
\caption{A heat-map showing the self-reported travelling patterns with different travelling companionship types from Europe participants}
\label{table:travel_experience_eu}
\centering
\pgfplotstabletypeset[%
     color cells={min=0,max=125,textcolor=black},
     /pgfplots/colormap={}{rgb255=(255,255,255) rgb255=(255,0,0)},
    /pgf/number format/fixed,
    /pgf/number format/precision=3,
    col sep=comma,
    columns/-/.style={reset styles,string type}%
]{
-,Alone, Colleagues, Friends, Family without kids, Family with kids
Once per week, 7, 1, 1, 4, 4
Once per month,28, 12, 22, 14, 13
Once per quarter,14, 17, 36, 25, 31
Once per half year,19, 20, 45, 37, 31
Once per year,26, 30, 50, 43, 36
Less than once per year,89,	82,	42,	64,	48
}
\end{table*}

\begin{table*}[!htb]
\caption{A heat-map showing the self-reported travelling patterns with different travelling companionship types from South Africa participants}
\label{table:travel_experience_sa}
\centering
\pgfplotstabletypeset[%
     color cells={min=0,max=50,textcolor=black},
     /pgfplots/colormap={}{rgb255=(255,255,255) rgb255=(255,0,0)},
    /pgf/number format/fixed,
    /pgf/number format/precision=3,
    col sep=comma,
    columns/-/.style={reset styles,string type}%
]{
-,Alone, Colleagues, Friends, Family without kids, Family with kids
Once per week, 13, 5, 9, 5, 9
Once per month,19, 8, 14, 11, 7
Once per quarter,18, 16, 28, 16, 16
Once per half year,6, 8, 16, 12, 21
Once per year,16, 16, 15, 18, 27
Less than once per year,19,	31,	10,	24,	10
}
\end{table*}

\end{document}